\documentclass{article}

\usepackage{arxiv}

\usepackage{amsmath}
\usepackage{comment}
\usepackage{bigints}
\usepackage{amssymb}
\usepackage[utf8]{inputenc} 
\usepackage{graphicx}
\usepackage{subcaption}
\usepackage[T1]{fontenc}    
\usepackage{url}            
\usepackage{booktabs}       
\usepackage{algorithm}
\usepackage{algorithmic}

\DeclareMathOperator*{\argmin}{argmin}
\usepackage{algorithm}
\usepackage{algorithmic}
\usepackage{amsfonts}       
\usepackage{nicefrac}       
\usepackage{microtype}      
\usepackage{lipsum}
\usepackage{hyperref}
\usepackage{mathtools} 
\usepackage{empheq} 
\newtheorem{theorem}{Theorem}[section]

\newtheorem{prop}[theorem]{Proposition}
\hypersetup{
    colorlinks=true,
    linkcolor=blue,
    filecolor=magenta,      
    urlcolor=cyan,
}

\title{Portfolio Stress Testing and Value at Risk (VaR) Incorporating Current Market Conditions}

\author{
  Krishan Mohan Nagpal  
  \thanks{Krishan Nagpal is a Managing Director in Corporate Risk in Wells Fargo \& Co. The opinions expressed here are those of the author and do not represent those of his employer Wells Fargo \& Co.}\\
  Wells Fargo \& Co. \\
  \texttt{krishan.nagpal@wellsfargo.com} \\
}

\begin{document}
\maketitle

\begin{abstract}
Value at Risk (VaR) and stress testing are two of the most widely used approaches in portfolio risk management to estimate potential market value losses under adverse market moves. VaR quantifies potential loss in value over a specified horizon (such as one day or ten days) at a desired confidence level (such as 95'th percentile). In scenario design and stress testing, the goal is to construct extreme market scenarios such as those involving severe recession or a specific event of concern (such as a rapid increase in rates or a geopolitical event), and quantify potential impact of such scenarios on the portfolio. Many approaches have been proposed for VaR and stress testing ranging from historical experience based to various statistical approaches. The goal of this paper is to propose an approach for incorporating prevailing market conditions in stress scenario design and estimation of VaR so that they provide more accurate and realistic insights about portfolio risk over the near term. The proposed approach is based on historical data where historical observations of market changes are given more weight if a certain period in history is "more similar" to the prevailing market conditions. Clusters of market conditions are identified using a Machine Learning approach called Variational Inference (VI) where for each cluster future changes in portfolio value are similar. VI based algorithm uses optimization techniques to obtain analytical approximations of the posterior probability density of cluster assignments (market regimes) and probabilities of different outcomes for changes in portfolio value. Covid related volatile period around the year $2020$ is used to illustrate the performance of the proposed approach and in particular show how VaR and stress scenarios adapt quickly to changing market conditions. In addition to estimation of VaR and stress scenario design, another advantage of the proposed approach is that classification of market conditions into clusters can provide useful insights about portfolio performance under different market conditions.



\end{abstract}


\section{Introduction}

VaR is a widely used measure in portfolio risk management as it provides a loss estimate over a specified time horizon for a desired confidence level. For example if ten day VaR for $97.5\%$ confidence level is $x$, then it implies that probability of portfolio loss exceeding $x$ over the future ten day period is $2.5\%$. There are some shortcomings of VaR as a risk measure with one main limitation being that it does not maintain the sub-additivity property (increased diversification within the portfolio may produce higher rather than lower VaR). Nevertheless VaR has become an important risk management measure and also plays an important role in computation of market risk capital for banks under Basel Committee on Banking Supervision
(BCBS) rules. Various approaches have been proposed for computation of VaR ranging from parametric to non-parametric distribution assumptions of risk factors, simulations based on historical changes or Monte Carlo simulation of risk factors based on some distributional assumptions (see for example \cite{Duffie}, \cite{Lins}, \cite{Jorion}, \cite{perignon} and \cite{Glasser} for overview of the different approaches). One of the main areas of focus in computation of VaR has been development of algorithms that realistically capture the heavy tailed distribution of portfolio returns often seen in the financial markets. 

Traditional approaches for VaR based on historical simulation may not adequately capture future volatility of the portfolio (see for example Perignon and Smith (\cite{perignon}). To capture future volatility of the portfolio over a short horizon more accurately it is also important to incorporate current market conditions (for example stress scenarios are likely to be less severe in benign markets compared to volatile markets). If risk factor distribution assumptions do not incorporate current market conditions and market trends, VaR is usually underestimated when market volatility suddenly increases and overestimated when the markets stabilize after a volatile period (this is also illustrated in the example described in the later section covering the Covid 2020 period). The goal of this paper is to propose an approach that incorporates current market conditions so that one captures near term portfolio risk in a more plausible way. To incorporate current market conditions in the proposed approach, we group market conditions into clusters where each cluster has a different pattern of market changes. Clusters incorporate information such as market trend/momentum and volatility that allows one to differentiate possible market changes and stress scenarios in near future. The proposed approach for computation of VaR combines clustering of market conditions with traditional historical simulation approach with one key difference. In the historical simulation based approach for VaR, one assumes that the historical observed risk factor changes used to obtain distribution of portfolio returns are equally likely while in the proposed approach here historical data is weighted differently and historical dates that are "similar" to the current market condition are given more weight. 

In stress tests, which typically are based on more severe market conditions than VaR, time horizon is not as specific as VaR and the goal is to understand impact on portfolio from an extreme but plausible changes in market conditions. The stress scenarios may try to quantify impact from macroeconomic scenarios (such as recession, impact of rapid increase in rates in 2022 and 2023 on regional banks and commercial real estate etc.), geopolitical risks or natural catastrophes. For financial institutions stress tests also play an important role in Federal Reserve’s capital adequacy tests (see for example Comprehensive Capital Analysis and Review 2016 \cite{CCAR}). One of the most common ways to create stress scenarios is to use some stressful financial period such as great recession of 2008 or Covid related stress period of 2020. However, historical stress scenarios may not be very useful if similar stress scenarios are unlikely to occur again in the near future. One of the primary challenges for such scenario designs is to create coherent and plausible stress shocks for several thousand risk factors (a bank portfolio can have well over hundred thousand risk factors that influence portfolio value) and especially so for rare specific events of concern for which the relevant historical data maybe very limited. For such scenarios where empirical data may be sparse, Bayesian network approaches proposed by Rebonato \cite{rebonato} can be used. In this approach transmission channels and conditional probabilities are defined for different events and though subjective, allow for interpretable scenario design incorporating subject matter expertise (in describing the linkages of events and conditional probabilities). The challenges in such approaches are in choosing reasonable conditional probabilities relevant for the specific scenario especially when dealing with large number of risk factors and limited relevant data.  

Stress scenario design differs from VaR estimation in one key way - in stress scenario design we are not only interested in quantifying potential loss in portfolio value but are also interested in determining plausible market conditions and risk factor changes that would result in stress loss.  The approach proposed here extends the approach described in Nagpal \cite{nagpal} of designing stress scenarios by synthesizing historical stress periods for the portfolio by linking stress scenarios to prevailing market conditions. Clustering approach used for stress scenario design is more granular than that used for VaR as clusters are based on not only severity of future losses but also changes in key risk factors so as to capture relationship between losses and key risk factor changes. Incorporating both loss severity and market changes in cluster identification allows one to determine, for example, whether stress scenario with increasing rates or decreasing rates is more likely in the near future. Utilizing historical data, clusters of market regimes are identified where in each cluster peak losses over the near future have both a) similar severity, and b) similar market changes in key risk factors. The proposed approach can also be customized to specific market conditions of concern such as stress scenario for increasing rate environment or high inflation periods. This is achieved by designing stress scenarios from only those historical periods that satisfy the market conditions of interest.

For both VaR and stress scenario design, the key step is identification of clusters or market regimes from historical data that best incorporate market trend/momentum and volatility in a manner that allows one to differentiate possible portfolio returns in the near future.  Variational Inference (VI) is used to identify clusters based on recent risk factor changes and portfolio trends so that for each cluster future portfolio return distributions are similar. For example in one cluster, portfolio value maybe more likely to increase while in another cluster the opposite may be the case. VI is generally used to obtain analytical approximations of the posterior probability density for graphical models where the observed data is dependent on unobserved latent variable. VI has been used in a wide range of applications to get approximate Bayesian posterior distributions conditioned on data (see for example (\cite{Bishop} and \cite{BleiStats}). Compared to other approaches based on sampling such as the Metropolis-Hastings algorithm (\cite{Gelfand}), VI can be more computationally efficient for large data sets or complex distributions. The concept of latent variables in creating posterior distributions in VI has a natural analog in financial data as market participants think of market conditions as "different regimes" where in each regime markets behave in a particular manner. For example market participants might think of market regimes as "rates increasing" vs. "rates decreasing" environments, or as "risk on" vs. "risk off" environments, where in each of such environments market changes are expected to follow a similar trend. In VI or other latent variable approaches, these market regimes such as "risk on" or "risk off" are not user specified but rather identified through analysis of the data. Nagpal \& Nagpal (\cite{udai}) have used a similar VI to propose an algorithms that can be applied to estimate near term market returns of macro indices such as S\&P. While that approach also provides a distribution of potential changes in portfolio value, it is not very suitable to estimate low probability extreme portfolio value changes in VaR and stress scenarios. 

In the proposed framework, portfolio changes or stress losses are classified into $m$ discrete categories such as large decline in portfolio value, unchanged, gains in portfolio value etc. In the case of stress scenario design, apart from portfolio value changes, these classifications may also include changes in risk factors of primary concern. For example if in stress scenario design one would like to create separate stress scenarios for increasing and decreasing rate environment, one would create categories of outcomes that involve both portfolio value changes as well as interest rate changes (an illustrative example is presented in the stress scenario design section). At any given time one is not sure about future portfolio change (which of these $m$ categories would occur) and the likelihood of these $m$ outcomes depends on the current market regime or cluster which is a latent variable. It is assumed that the underlying market data is a mixture model of Gaussian clusters. Clusters are based on risk factors and portfolio performance such as portfolio value or interest rate changes in most recent $n$ days, levels of volatility measures such as VIX etc. From each cluster or market regime probability of future portfolio changes into the $m$ discrete outcomes follows a Dirichlet distribution. Similar mixture models have been considered in \cite{dele} and for dynamical systems in \cite{pineda} and in both cases parameter estimation is based on Expectation-Maximization. Variational approaches for mixtures of linear mixed models have also been considered \cite{tan}. The main advantages of using the proposed approach in financial forecasting and risk management are a) ease of implementation due to relatively easy optimization problem, b) one obtains not only an estimate but the also distribution of outcomes which can be useful in portfolio risk management, and c) by identifying clusters, one gains insight about market conditions in which future changes follow similar patterns.

This paper is organized as follows. The next section contains a brief introduction of variational inference and mean-field family of distributions. The underlying assumptions and modeling framework for VaR are described in Section 3. Section 4 describes the framework and assumptions for stress scenario design. Illustrative examples to demonstrate the performance of the proposed approaches in both normal and volatile market conditions are included in the corresponding sections. Section 5 concludes with a summary and the Appendix contains the derivation of the variational parameter estimates.

\section{Brief Introduction to Variational Inference}

In Bayesian statistics the goal is to infer the posterior distribution of unknown quantities using observations. For some complex problems it is assumed that observations, denoted by $x=\{x_1,\cdots,x_N\}$ in this section, are linked to latent variables $z=\{z_1,\cdots,z_K\}$ which are drawn from prior distribution $p(z)$. The likelihood of observation $x$ depends on $z$ through the distribution $p(x|z)$. In Bayesian learning and estimation problems the goal is to learn the posterior distribution $p(z|x)$ of latent variables conditioned on the data.

VI is a widely-used method in Bayesian machine learning to approximate posterior distributions. In VI a suitable family of distributions $\mathcal(F)$ is chosen which is complex enough to capture the attributes of the data and yet simple enough to be computationally tractable. The best distribution is obtained from an optimization to minimize Kullback-Leibler (KL) divergence with the exact posterior (see for example (\cite{Bishop} and \cite{BleiStats}): 
\begin{equation}
\label{q*def}
    q^*(z)=\argmin_{q(z) \in \mathcal(F)} \; KL \left( q(z) || p(z|x) \right)
\end{equation}
where the KL divergence, a measure of information-theoretic distance between two distributions, is obtained as follows where $\mathbb{E}_q$ denotes expectation with respect to $q(z)$:
\begin{align}
    KL \left( q(z) || p(z|x) \right) & :=  \mathbb{E}_q \left[\text{log} \ q(z)\right] - \mathbb{E}_q \left[ \text{log} \ p(z|x)\right] \nonumber \\
    \label{KL}
    & = \mathbb{E}_q \left[ \text{log} \ q(z)\right] - \mathbb{E}_q \left[ \text{log} \ p(z,x)\right] + \text{log} \ p(x)
\end{align}
Since log $p(x)$ does not depend on the distribution $q(z)$, the optimal distribution $q^*(z)$ that minimizes KL divergence in (\ref{q*def}) also maximizes the Evidence Lower Bound (ELBO) defined as follows:
\begin{equation}
\label{elbo}
    \text{ELBO}(q):=\mathbb{E}_q \left[ \text{log} \ p(z,x)\right] -\mathbb{E}_q \left[ \text{log} \ q(z)\right]
\end{equation}

Here we will work with the mean-field variational family of distributions $q(z)$ in which one assumes that the latent variables are mutually independent and probability density of each latent variable is governed by distinct factors (see \cite{Bishop} and \cite{BleiStats} for more details). This assumption greatly simplifies the computation of the optimal parameters of the variational distribution $q(z)$. Due to the assumption of mutual independence, if there are $K$ latent variables $z=\{z_1, \hdots, z_K\}$,  the joint density function is of the form
\[
q(z)= \prod_{k=1}^K q_k(z_k)
\]
where $q_k(z_k)$ is the density of the k'th latent variable $z_k$ and each $q_k$ has its own set of parameters. For the mean-field variational family of distributions, the most commonly used approach for obtaining parameters that maximize the ELBO (defined in (\ref{elbo})) is coordinate ascent variational inference (CAVI). In this approach, which has similarities to Gibbs sampling, one optimizes parameters for each latent variable one at a time while holding others fixed, resulting in a monotonic increase in the ELBO (\cite{Bishop},\cite{BleiStats}). 

It can be shown (\cite{Bishop}) that for the mean-field variational family, $q_k^*(z_k)$, the optimal distribution for $q_k(z_k)$ that maximizes ELBO (\ref{elbo}), satisfies the following: 
\[
    q_k^*(z_k)\propto \text{exp} \left\{ \mathbb{E}_{-k} \left[ \text{log} \ p(z_k \mid z_{-k}, x)\right] \right\}
\]

where $z_{-k}$ denotes all $z_i$ other than $z_k$ and $\mathbb{E}_{-k}$ represents expectation with respect to all $z_i$ other than $z_k$. Thus the optimal variational distribution of latent variable $z_k$ is proportional to the exponentiated expected log of the posterior conditional distribution of of $z_k$ given all other latent variables and all the observations. Since $p(z_{-k}, x)$ does not depend on $z_k$ (due to independence of latent variables $z_i$), one can write the above equivalently as
\begin{equation}
\label{meanfield}
    q_k^*(z_k)\propto \text{exp} \left\{ \mathbb{E}_{-k} \left[ \text{log} \ p(z_k , z_{-k}, x)\right] \right\}
\end{equation}
For example, if there are three latent variables $z_1,z_2$ and $z_3$, the above implies
\[
q_1^*(z_1)\propto \text{exp} \left\{ \mathbb{E}_{z_2,z_3} \left[ \text{log} \ p(z_1, z_2, z_3, x)\right] \right\}
\]
The proposed approach is based on the mean-field family and thus (\ref{meanfield}) will be instrumental in obtaining the parameters of the posterior distribution.

\section{VaR Incorporating Current Market Conditions}

In the following section we will first describe the framework of modeling the market returns. 

\subsection{Underlying Model for Data Generation}

If $D$-Day VaR on any given day for confidence level $\alpha$ (for example $95\%$ or $99\%$) is $y$ then it implies that probability of portfolio loss exceeding $y$ over the future $D$ days is $(1-\alpha)$. Thus for computation of $D$-Day VaR ($D$ typically ranges from $1$ to $10$ days) one has to first obtain the distribution of portfolio value change over the next $D$ days. 

\begin{figure}
        \includegraphics[width=\textwidth]{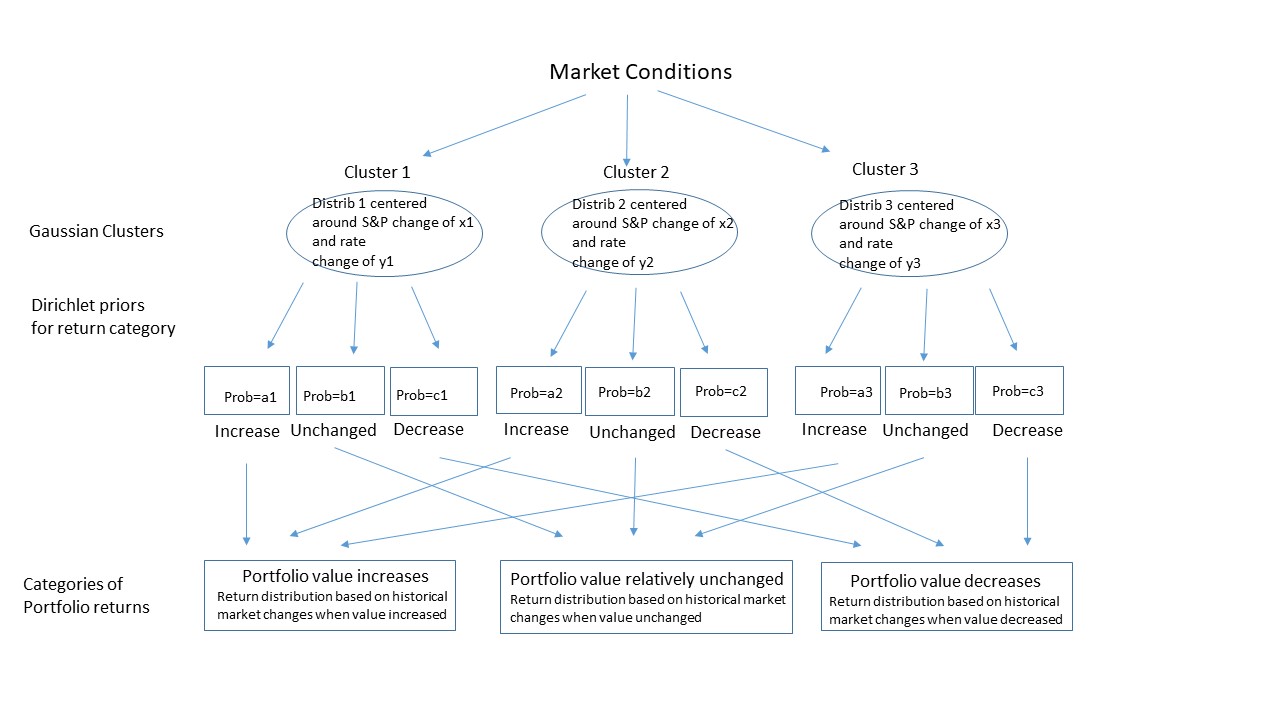}
        \caption{Illustrative example of modeling framework based on three clusters ($K=3$) and three categories of portfolio value changes from each cluster ($J=3$). Variational Inference is used to identify clusters and estimate parameters of Dirichlet distribution for different categories of outcomes (portfolio value changes) from each cluster. Distribution of portfolio returns within each of the $J$ categories is obtained from historical data of risk factor changes.}
    \label{illustration}
 \end{figure}

Before describing mathematical details, we describe the assumed underlying model for portfolio return over the next $D$ days. The modeling approach is not based on a time series model but rather based on patterns linking portfolio performance to macro market data, volatilities and their trends. This model calibrated to historical data is used to generate distribution of future portfolio returns from which one obtains VaR. The approach is illustrated by Figure \ref{illustration}. Each day it is assumed that the market is in one of $K$ possible clusters which are hidden/latent variables. Clusters, which are identified by recent changes in key risk factors (such as changes in rates, credit spreads or stock indices) or portfolio value, carry the information about future potential changes in market value of the portfolio. Different clusters have different distribution of potential outcomes - for example portfolio value may be more likely to increase than decrease from cluster A while the opposite maybe the case for cluster B. Clusters depend on market data $x_t \in \mathbb{R}^n$ which is known at time $t$. The elements of input $x_t$ are chosen to best differentiate clusters and could be variables such as S\&P or interest rate changes over the most recent $l$ days or their volatilities or some measures of market activity such as trading volumes. Choice of best input variables to identify clusters is of course very important but will not be discussed in this paper - we will assume that we have already identified the choice of $n$ input variables in $x_t$ that are most useful in differentiating clusters. The optimal choice of components of $x_t$ could be based on choosing the combination that has the best out of sample performance using the proposed algorithm. For each $t$, $c_t \in \mathbb{R}^K$ will denote the indicator vector which describes which cluster $x_t$ belongs to. The indicator vector $c_t$ has $K-1$ elements equal to $0$ and one element equal to $1$ that corresponds to the cluster assignment of $x_t$. For example, if $x_t$ is in the second cluster then 
\[
c_t= \begin{bmatrix}
    0  \\ 1 \\ 0 \\ \vdots \end{bmatrix}
\]
We will denote $c_{t(k)}$ as the $k'th$ element of $c_t$ and as noted above $c_{t(k)}=1$ for only one $k \in \{1, \cdots, K \}$ and is zero for all other $k$. For example if the vector $c_t$ is as above, $c_{t(2)}=1$ and $c_{t(k)}=0$ for all $k \neq 2$. 

For computational tractability, a number of assumptions are made about clusters. 
We will assume that the mean and variance of clusters do not change over time and the prior probability for the state $x_t$ to be in cluster $k$ at any given time is $\pi_k$ (the sum of $\pi_k$ over $K$ has to be one as at every time $x_t$ is in one of the $K$ clusters). The mean of $x_t$ for cluster $k$, which we denote by $\mu_k$, is not known. It is assumed that for each cluster $k$, $\mu_k$ is normally distributed with mean $\mu_{k0}$ and variance $R_{k0}$. Within each cluster $k$, $x_t$ is normally distributed around the cluster mean $\mu_k$ with variance $M$, which is the same for all clusters. It is assumed that explanatory variables $x_t$ have been suitably adjusted for mean and drift terms so that the cluster mean $\mu_k$ does not depend on time $t$. For example, transforming market indices with drift (such as S\&P) to relative changes, removes the drift and makes the data almost stationary.


Future changes in the portfolio value ($D$-Day P\&L in case of VaR) from each cluster are assumed to fall into $J$ possible categories of outcomes ranging from large increase in value to large decrease in the value of the portfolio. The proportions of portfolio changes from a cluster $k$ to the $J$ categories will be denoted by $\theta_k$ (a $J$ dimensional vector of non-negative numbers that sum to one, i.e. $\sum_{j=1}^J \theta_{k_j}=1$ where $\theta_{k_j}$ is the probability of the $j'th$ category outcome from cluster $k$). The proportions $\theta_k$ are not known but have priors of Dirichlet distribution of order $J$. The $J$ categories of portfolio change are the same for all the clusters but the probabilities of the $J$ outcomes varies from cluster to cluster. For example from cluster $1$, categories corresponding to portfolio value increase may be more likely than categories corresponding to portfolio value decrease while the opposite may be the case for cluster $2$. Similar to cluster assignment variable $c_t$, for each $t$ the variable $d_t\in \mathbb{R}^J$ assigns the $D$ day P\&L from time $t$ to $t+D$ into one of the $J$ categories. The indicator vector $d_t$ has $J-1$ elements equal to $0$ and one element equal to $1$ that corresponds to the $D$ day P\&L category starting from time $t$. For example, if $D$ day P\&L from time $t$ is in the third category then
\[
d_t= \begin{bmatrix}
    0  \\ 0 \\ 1 \\ 0 \\ \vdots \end{bmatrix}
\]
We will denote $d_{t(j)}$ as the $j'th$ element of $d_t$ and as noted above $d_{t(j)}=1$ for only one $j \in \{1, \cdots, J \}$ and is zero for all other $j$. For example if the vector $d_t$ is as above, $d_{t(3)}=1$ and $d_{t(j)}=0$ for all $j \neq 3$.
Note that since $d_t$ describes the future $D$ day P\&L change bucket at time $t$, it is not known until time $t+D$. The goal of the model is to have a good estimate of $d_t$ based on information available at time $t$. 

For portfolio value change within each of the $J$ categories of possible outcomes (such as decrease in portfolio value), we will assume Empirical Distribution (ED) derived from historical data which is described in greater details later. In other words, the empirical distribution of portfolio value change within each category ($\text{Prob}(\text{$D$ day P\&L =$x$ | $d_{t(j)}=1$} )$) is derived from past historical data of $D$ day P\&L from periods when the portfolio change was in category $j$. Utilizing this, one can obtain the distribution of portfolio value changes over $D$ days as follows:


\begin{align}
\text{Prob} \left( \text{$D$ day P\&L at $t$ is $x$} \right) & = \sum_{j=1}^J \text{Prob} \left( \text{$D$ day P\&L} = x \mid  
d_{t(j)}=1 \right) *\text{Prob}\left( d_{t(j)}=1 \right) \nonumber \\
\label{probpnlcat}
& = \sum_{j=1}^J \  \text{Prob} \left( \text{$D$ day P\&L} = x \mid  d_{t(j)}=1 \right) 
\left[ \sum_{k=1}^K \text{Prob} \left( d_{t(j)}=1 \mid c_{t(k)}=1 \right) \text{Prob}\left(c_{t(k)}=1 \right)  \right]
\end{align}

Apart from empirical historical data based $\text{Prob} \left( \text{$D$ day P\&L} = x \mid  d_{t(j)}=1 \right)$, all other probabilities on the right side of equation above are obtained using VI. In particular VI calibration at time $t$ is used to determine a) $\text{Prob}(\text{$c_{l(k)}=1$})$ which are probabilities of cluster assignments for all $x_l$ with $t-T \leq l \leq t$ where $T$ is the number of past historical dates used in computation of VaR, and b) $\text{Prob}(\text{$d_{t(j)}=1  \mid  c_{t(k)}=1$})$ which are the probabilities of $J$ category outcomes from each of the cluster. Once we have the distribution of portfolio value change over the next $D$ days (from equation (\ref{probpnlcat})), one can obtain VaR for desired confidence level.

Before describing how one obtains empirical distribution ED for each of the $J$ categories based on historical data, we briefly review the well known historical simulation approach for VaR. Recall that for computation of $D$-Day VaR the first step is obtain the distribution of portfolio value change over $D$ days.  If VaR is based on observed market changes over the most recent $T$ time periods (for example if VaR is based on the past one year of observations then $T$ is approximately $250$ daily observations from the preceding one year), the distribution of portfolio value change over $D$ days is derived from those $T$ historical observations of market changes. In traditional historical simulation approaches, certain assumptions are made such as each of the daily observations are equally likely. The computation of portfolio value change for specified market risk factor changes could be based on Taylor series approximation of portfolio Greeks or full valuation of all the positions. At time $t$ (we will assume $t$ is the current and most recent time for which data is available) let the $D$ day simulated portfolio value change over the previous $T$ observations of market changes over $D$ days be $Hist\_Ret_{t}$: : 
\[
Hist\_Ret_{t}=\begin{bmatrix} P\&L_{t,1} , & \cdots , & P\&L_{t,T} \end{bmatrix} 
\]
where
\[P\&L_{t,i}:=\text{Change in value of time $t$ portfolio for market changes from day $(t-D-i)$ to day $(t-i)$} 
\]
Note that $P\&L_{t,i}$ is a simulated value of portfolio return for the portfolio at time $t$ based on market changes from day $(t-D-i)$ to day $(t-i)$. Thus $P\&L_{t,i}$ may be quite different from actual portfolio value change from day $(t-D-i)$ to day $(t-i)$ since the portfolio at time $t$ could be different compared to what it was at time $(t-D-i)$. Since the past historical market data of risk factors that influence portfolio value is known, the vector of simulated historical portfolio returns $Hist\_Ret_{t}$ is known at time $t$. 

In the proposed approach, empirical P\&L distribution for each of the $J$ categories of outcomes is obtained by dividing daily outcomes from historical simulation into $J$ non overlapping categories of outcomes. Category $j$ of returns would have portfolio value changes between $P\&L^j_{min}$ and $P\&L^j_{max}$. To ensure all possible outcomes fall into only one of the $J$ buckets we will assume the boundaries of returns have been chosen so that $P\&L^1_{min} < P\&L^1_{max}=P\&L^2_{min} < P\&L^2_{max} \hdots P\&L^J_{min} < P\&L^J_{max}$. Let $Hist\_Ret_{t}^j$ denote the set of portfolio value changes for category $j$ and let $n_{t,j}$ be the number of elements $Hist\_Ret_{t}$ that are in $Hist\_Ret_{t}^j$. For portfolio return distribution in category $j$, we will assume that portfolio return can be only one of the elements of vector $Hist\_Ret_{t}^j$ and each of those outcomes is equally likely (within portfolio change in category $j$, each element of $Hist\_Ret_{t}^j$ has probability $\frac{1}{n_{t,j}}$). These are summarized below: 
\begin{align}
    \label{jcategory}
Hist\_Ret_{t}^j=\begin{bmatrix} \text{$P\&L_{t,i}$ if } P\&L_{t,i} \in Hist\_Ret_{t} \text{ and } P\&L^j_{min} \leq P\&L_{t,i} \leq P\&L^j_{max} \end{bmatrix} \\
n_{t,j} = |Hist\_Ret_{t}^j|=\text{ the number of elements in } Hist\_Ret_{t}^j \; ; \; (\text{note that } \sum_{j=1}^J n_{t,j}=T )  \nonumber \\
\text{Probability ($D$ day P\&L is P$\&L_{t,i} \ \mid$ \ category $j$)}= \frac{1}{n_{t,j}} \; \; \text{if } P\&L_{t,i} \in Hist\_Ret_{t}^j \text{ and $0$ otherwise} \nonumber
\end{align}

We will make a simplifying assumption that each observation of input-output data is independent ($(x_t,c_t,d_t)$ is independent of $(x_\tau,c_\tau,d_\tau)$ whenever $t\neq \tau$). In case of daily prediction, this assumption implies that each day provides a new independent observation. This is a simplifying assumption as there may be some overlap in market data $x_t$ and $x_{t+1}$ (as they may incorporate data from same prior days). It is possible to to drop the independence assumption and extend the VI framework to the case where $x_t$ is the state of a linear dynamical system as in \cite{pineda}. However, such an approach leads to greater complexity as one must also estimate the parameters that describe the time evolution of the system.

\clearpage
The assumptions described above imply that the observed data (input $x_t$ and output which is the next $D$-day P\&L) is generated by the following model

\begin{center}
\fbox{%
    \parbox{0.8\linewidth}{
        \textbf{Hyperparameters and prior information for computation of $D$-Day P\&L}
        \begin{itemize}
            \item $J$ is the number of categories of future $D$-Day portfolio changes. The lower bound and upper bound of portfolio change for category $j$ are $P\&L^j_{min}$ and $P\&L^j_{max}$.
            \item Within each category $j$, portfolio value change over $D$ days is one one of the elements of $Hist\_Ret_{t}^j$ (described in (\ref{jcategory})) with each element equally likely with probability $\frac{1}{n_{t,j}}$.
            \item $K$ is the number of clusters of input space $x_t$ (where $x_t \in \mathbb{R}^{n}$). 
            \item The fraction of times $x_t$ is in cluster $K$ is $\pi_k$. Thus the sum of $\pi_k$ over all $k$ is one.
            \item For cluster $k$, the mean of $x_t$ is $\mu_k$, which is normally distributed with mean $\mu_{k0}$ and covariance $R_{k0}$. In other words, for cluster $k$, $\mu_k \sim \mathcal{N}(\mu_{k0},R_{k0})$. 
            \item If $x_t$ is in cluster $k$ then $x_t$ is normally distributed with  mean $\mu_{k}$ and covariance $M$. In other words if $c_{t(k)}=1$ then $x_t \sim \mathcal{N}(\mu_{k},M)$. Note that while each cluster has a different mean $\mu_k$, we are assuming the the same variance $M$ for $x_t$ in each cluster.
            \item For cluster $k$, the proportions vector $\theta_k$ describes proportion of times $D$ day portfolio return is in each of the $J$ categories (note $\sum_{j=1}^J \theta_{k_j}=1$). The prior for proportions $\theta_k$ for cluster $k$ is $J$-Dirichlet distribution with parameters $\alpha_k$. 
        \end{itemize}
        \textbf{Generative process for $D$-Day portfolio value change}
        \begin{enumerate}
            \item For $\; k \in \left[ 1, \hdots, K \right]$, draw $\mu_k \ \sim \mathcal{N}(\mu_{k0},R_{k0})$
            \item For $k \in \left[ 1, \hdots, K \right]$, draw proportions $\theta_k \sim Dirichlet(\alpha_k)$ (note $\sum_1^J \theta_{k_j}=1$)
            \item For $t \in \left[ 1, \hdots T \right]$   (VaR is based on most recent $T$ daily observations)
            \begin{enumerate}
                \item Draw latent variable $c_{t(k)} \ \sim \ \text{Cat} (\pi_k)$ which describes the cluster assignment for time $t$. $c_{t(k)}$ is an indicator variable which is 1 for only one of its $K$ elements and zero for all other $K-1$ elements.
                \item If the cluster assignment in previous step is $k$, draw input $x_t  \ \sim \mathcal{N}(\mu_{k},M)$
                \item If the cluster assignment is $k$, draw future $D$ day portfolio change category assignment $d_t \ | \theta_k \sim \text{Cat}(\theta_k)$. ($d_{t(j)}$ is $1$ for only one of the $J$ elements of $d_t$).
                \item If the portfolio change is in category $j$, draw randomly one of the $n_{t,j}$ portfolio returns from $Hist\_Ret_{t}^j$ (within category $j$, all returns in the set $Hist\_Ret_{t}^j$ are equally likely).
            \end{enumerate}
        \end{enumerate}
    }
}
\end{center}

\clearpage

\subsection{Variational Inference Calibration for VaR}


We now summarize the above prior information where $\pi_k$, $\mu_{k0}$, $R_{k0}$ and $\alpha_k$ for $k \in \{1, \cdots, K\}$ and $j \in \{1, \cdots, J\}$ are hyperparameters for cluster and category assignments:
\begin{align}
\label{cluster}
p(c_{t(k)}=1) & = \pi_k \; \; \text{for  k } \in \{1,\cdots,K\} \; \text{where } \sum_1^K \pi_k=1 \\
\label{dirich}
p(\theta_k ; \alpha_k) & =  \frac{\Gamma (\sum_{j=1}^J \alpha_{k_j})}{\prod_{j=1}^J \Gamma (\alpha_{k_j})} \prod_{j=1}^J \theta_{k_j}^{\alpha_{k_j} -1} \; \; \text{where } \theta_k = \begin{bmatrix}
    \theta_{k_1}  \\ \vdots \\ \theta_{k_J} \end{bmatrix} \; 
    \text{and } \alpha_k = \begin{bmatrix}
   \alpha_{k_1}  \\ \vdots \\\alpha_{k_J} \end{bmatrix} \\
\label{muprior}
    p(\mu_k ; \mu_{k0},R_{k0}) & =  \frac{\text{exp}(-\frac{1}{2}(\mu_k-\mu_{k0})'R_{k0}^{-1}(\mu_k-\mu_{k0}))}{\sqrt{(2 \pi)^n|R_{k0}|}} \; \; \; \text{for  k } \in \{1,\cdots,K\} \\
    \label{xcluster}
    p(x_t | \mu_k, c_{t(k)}=1) & =\frac{\text{exp}(-\frac{1}{2}(x_t-\mu_{k})'M^{-1}(x_t-\mu_{k}))}{\sqrt{(2 \pi)^n|M|}}
 \end{align}
 
In computation of VaR, new VI parameters will be estimated everyday based on the most recent $T$ observations. Recall that while $x_t$ is known at time $t$, $d_t$ which describes the category of future $D$ days P\&L from time $t$ is known only on day $t+D$. Since both input and output are needed for model calibration, at time $t$ the latest daily observation of both input and output is for time $t-D$. For ease of notation we will assume that we are at time $T+D$ and the $T$ observations used for model calibration are $i \in \left[ 1, \hdots,  T \right]$ and the the VI parameters are estimated from these $T$ observations. The following notation will be used to denote the vectors:
\begin{equation}
\text{\bf{c}}=\{c_1, \hdots, c_T \}\ , \text{\bf{d}}=\{d_1, \hdots, d_T \}\ , \ \text{\bf{x}}=\{x_1, \hdots, x_T \}\ , \ \boldsymbol\mu =\{ \mu_1, \hdots, \mu_K \} \ , \ \boldsymbol\theta =\{ \theta_1, \hdots, \theta_K \} \
\end{equation}

Recall that the indicator function $c_{t(k)}=1$ if at time $x_t$ is in cluster $k$ at time $t$ and $0$ otherwise. Similarly the indicator function $d_{t(j)}=1$ if portfolio value change for the next $D$ days is in category $j$. Given these indicator functions, one notes that:
\begin{align}
\label{xprior}
    p(x_t | c_{t}, \boldsymbol\mu) & =  \prod_{k=1}^K p(x_t | \mu_k, c_{t(k)}=1)^{c_{t(k)}}  \\
 \label{dprior} 
 p(d_{t(j)}=1 | c_{t}, \boldsymbol\theta)  & = \prod_{k=1}^K \theta_{k_j}^{c_{t(k)}} \; \; \; (\text{since } p(d_{t(j)}=1 | c_{t(k)}=1, \theta_k)=\theta_{k_j})
\end{align}

With the data generating process described above one observes that 
\begin{equation}
\label{factorization}
    p( \boldsymbol\mu ,\boldsymbol\theta, \text{\bf{c}}, \text{\bf{d}}, \text{\bf{x}})= \prod_{k=1}^K p(\mu_k) \prod_{k=1}^K p(\theta_k) \prod_{t=1}^T \left[ p(c_t) p(x_t|\boldsymbol\mu, c_t) p(d_t|c_t, \boldsymbol\theta)  \right]
\end{equation}
The above factorization implies
\begin{align}
\label{logP}
\text{log }  p( \boldsymbol\mu ,\boldsymbol\theta, \text{\bf{c}}, \text{\bf{d}}, \text{\bf{x}})
& =  \sum_{k=1}^K \text{log }  p(\mu_k) +\sum_{k=1}^K \text{log }  p(\theta_k)\nonumber \\
& + \sum_{t=1}^T \left[ \text{log }  p(c_t) + \text{log }  p(x_t \mid \boldsymbol\mu, c_t) + \text{log }  p(d_{t} \mid c_t, \boldsymbol\theta) \right]
\end{align}

For approximating the posterior probability density of latent variables $p(\boldsymbol\mu , \boldsymbol\theta, \text{\bf{c}}  \mid \text{\bf{x}}, \text{\bf{d}})$ by variational distribution $q(\boldsymbol\mu , \boldsymbol\theta, \text{\bf{c}})$, we will consider the following mean-field distribution where $\mu$, $\theta$ and $c$ are independent and governed by their own variational parameters (with the variational parameters being $\hat{\alpha}_k, \hat{\mu}_k, \hat{R}_k$ and $\phi_{t}$) :
\begin{equation}
\label{meanfield1}
    q(\boldsymbol\mu , \boldsymbol\theta, \text{\bf{c}})= q_\mu (\boldsymbol\mu)q_\theta (\boldsymbol\theta)q_c (\text{\bf{c}}) =  \prod_{k=1}^K q_\mu (\mu_k; \hat{\mu}_k, \hat{R}_k) \prod_{k=1}^K q_\theta (\theta_k ; \hat{\alpha}_k) \prod_{t=1}^T  q_c (c_t ; \phi_t)  
\end{equation}
Please note that the superscript " $\hat{ }$ " will denote variational parameters of the corresponding distribution. The mean-field family of distributions above may not contain the true posterior because of the assumption that $\boldsymbol\mu, \boldsymbol\theta$ and $\text{\bf{c}}$ are independent. The independence assumption leads to much more tractable estimation of the the posterior probability density of the latent variables. The distributions $q_c$ in terms of its variational parameter $\phi_{t}$ are assumed to be as follows:

\begin{equation}
   \label{qc1}
q_c(c_{t(k)}=1)  =\phi_{t_k} \; \; \; \text{for  t } \in \{1,\cdots,T\} \;  \text{and for  k } \in \{1,\cdots,K\} 
\end{equation}

Without making any assumptions on the family of distributions for $q_\mu$ and $q_\theta$, it is shown in the Appendix that these posteriors distributions are of the same form as priors as described below in terms of their variational parameters:   
\begin{align}
\label{qmu1}
q_\mu(\mu_k ; \hat{\mu}_k, \hat{R}_k) & =  \frac{\text{exp}(-\frac{1}{2}(\mu_k-\hat{\mu}_k)'\hat{R}_k^{-1}(\mu_k-\hat{\mu}_k))}{\sqrt{(2 \pi)^n|\hat{R}_{k}|}} \; \; \; \text{for  k } \in \{1,\cdots,K\} \\
\label{qtheta1}
q_\theta(\theta_k ; \hat{\alpha}_k) & =  \frac{\Gamma (\sum_{j=1}^J \hat{\alpha}_{k_j})}{\prod_{j=1}^J \Gamma (\hat{\alpha}_{k_j})} \prod_{j=1}^J \theta_{k_j}^{\hat{\alpha}_{k_j} -1} \; \; \; \text{for  k } \in \{1,\cdots,K\} \; \ \text{where } \; \  \hat{\alpha}_k := \begin{bmatrix}
   \hat{\alpha}_{k_1}  \\ \vdots \\ \hat{\alpha}_{k_J} \end{bmatrix}
\end{align}
where all the variational parameters $\hat{\mu}_k, \hat{R}_k$ and $\hat{\alpha}_{k}$ are described below in equations (\ref{phitk}) to (\ref{rtk1}). From proportions distribution (\ref{qtheta1}) one notes that variational distribution of a $j'th$ category P\&L outcome from cluster $k$ can be obtained in terms of variational parameters $\hat{\alpha}_k$ as
\begin{equation}
\label{catprobvar}
    \mathbb{E}_{q_\theta}(d_{t(j)}=1|c_{t(k)}=1)= \ \mathbb{E}_{q_\theta}(\theta_{k_j})= \ \frac{\hat{\alpha}_{k_j}}{\sum_{i=1}^J\hat{\alpha}_{k_i}}
\end{equation}

The variational posterior distribution $q(\boldsymbol\mu, \boldsymbol\theta , \bf{c})$, which approximates  $p(\boldsymbol\mu, \boldsymbol\theta , \bf{c} | \bf{x}, \bf{d})$ is obtained by maximizing ELBO (\ref{elbo}). The variational parameters $\{\hat{\mu}_k, \hat{R}_k, \hat{\alpha}_k, \phi_{t_k}\}$ have the following interpretation for the posterior distribution: $\phi_{t_k}$ is the probability of the $k'th$ cluster assignment for data $x_t$; parameters $\hat{\mu}_k, \hat{R}_k$ are the mean and variance of the $k'th$ cluster of input $x_t$; and $\hat{\alpha}_k$ are the Dirichlet distribution parameters for the posterior distribution of $\theta_k$.  Utilizing (\ref{logP}) and (\ref{meanfield1}), ELBO function (\ref{elbo}) can be expressed as follows where the expectation is with respect to variational distribution $q(\boldsymbol\beta, \boldsymbol\mu , \bf{c})$: 
\begin{align}
\label{elbo1}
    \text{ELBO}(\hat{\mu}_k, \hat{R}_k, \hat{\alpha}_k, \phi_{t_k}) & =  \mathbb{E}_q \left[ \text{log }  p( \boldsymbol\mu ,\boldsymbol\theta, \text{\bf{c}}, \text{\bf{d}}, \text{\bf{x}}) \right]
    - \mathbb{E}_q \left[ \text{log } q(\boldsymbol\mu , \boldsymbol\theta, \text{\bf{c}}) \right]
    \nonumber \\
   & = \sum_{k=1}^K \mathbb{E}_q \left[ \text{log} \ p(\mu_k) \right] + \sum_{k=1}^K \mathbb{E}_q \left[ \text{log} \ p(\theta_k) \right] \nonumber \\
    & + \sum_{t=1}^T \Big( \mathbb{E}_q \left[ \text{log} \ p(c_t) \right] + \mathbb{E}_q \left[ \text{log} \ p(x_t |\boldsymbol\mu, c_t) \right] + \mathbb{E}_q \left[ \text{log} \ p(d_{t} |c_t, \boldsymbol\theta) \right] \Big) \nonumber \\
    & - \sum_{t=1}^T \mathbb{E}_q \left[ \text{log} \ q_c(c_t) \right] - \sum_{k=1}^K \mathbb{E}_q \left[ \text{log} \ q_\mu(\mu_k) \right] - \sum_{k=1}^K \mathbb{E}_q \left[ \text{log} \ q_\theta (\theta_k) \right] 
\end{align}

The variational parameters $\{\hat{\mu}_k, \hat{R}_k, \hat{\alpha}_k, \phi_{t_k}\}$ are to be estimated so that the ELBO function described above is maximized.


The variational parameters that maximize the ELBO function in (\ref{elbo1}) can now be obtained based on (\ref{meanfield}). It is shown in the Appendix that the optimal variational parameters satisfy the following equations:
\begin{align}
\label{phitk}
    \phi_{t_k} & = \frac{r_{t(k)}}{\sum_{k=1}^K r_{t(k)}} \; \text{ for } k \in \{1, \hdots,K\} \; \text{and } t \in \{1, \hdots,T\} \\
    \hat{R}_{k} & =  \left[R_{k0}^{-1}  + M^{-1} \sum_{t=1}^T \phi_{t_k} \right]^{-1} \\
\hat{\mu}_{k} & =  \hat{R}_{k} \left[ R_{k0}^{-1} \mu_{k0} + M^{-1} \sum_{t=1}^T \phi_{t_k} x_t       \right] \\
\label{alphaest}
\hat{\alpha}_{k_j}  & = \left[ \alpha_{k_j} + \sum_{t=1}^T \phi_{tk} \ d_{t(j)} \right] \; \;  \text{for } j\in \{1,\cdots,J\} \; \text{and } k \in \{1,\cdots,K\} \\
 & \text{ (note that $d_{t(j)}$ is an indicator function for the category at time $t$ that is $1$ only for one $j$ and $0$ otherwise)}  \nonumber
\end{align}
where $r_{t(k)}$ is defined as follows in which $\Psi(.)$ is the digamma function (which is the logarithmic derivative of the Gamma function):
\begin{equation}
\label{rtk1}
    r_{t(k)}:=\text{exp} \left[ \text{log} (\pi_k) + x_t'M^{-1} \hat{\mu}_k - \frac{1}{2} \text{trace}(M^{-1} (\hat{\mu}_k \hat{\mu}_k'+\hat{R}_k))
    + \Psi (\sum_{j=1}^J \hat{\alpha}_{k_{j}}d_{t(j)} ) - \Psi(\sum_{j=1}^J\hat{\alpha}_{k_j})  \right]
\end{equation}

Note that since the indicator function $d_{t(j)}$ is $1$ for only one $j$, $\Psi (\sum_{j=1}^J \hat{\alpha}_{k_{j}}d_{t(j)} )=\Psi ( \hat{\alpha}_{k_{j*}} )$ where $d_{t(j*)}=1$. 

One of the most commonly used algorithms for obtaining optimal variational parameters is Coordinate Ascent Variational Inference (CAVI) (see for example \cite{BleiStats}). Here parameters are updated one at a time, keeping other parameters constant. The process is repeated until the ELBO converges. The CAVI algorithm for estimating parameters is described in Algorithm \ref{cavialg}. The algorithm may however converge to a local maximum and thus running the algorithm with different initial estimates of the variational parameters can improve the approximated model posterior.

\begin{algorithm}
\caption{CAVI algorithm for estimating variational parameters} \label{cavialg}
\renewcommand{\algorithmicrequire}{\textbf{Input:}}
\renewcommand{\algorithmicensure}{\textbf{Output:}}
\begin{algorithmic}
\REQUIRE Data $x_t$ and $d_{t}$ for $t \in \{1, \cdots,T\}$ 
\REQUIRE Hyperparameters: Number of clusters $K$; intra-cluster variance M; $\pi_k$, $\mu_{k0}$ and $R_{k0}$ for $k \in \{1, \cdots, K\}$ 
\STATE{}
\STATE{Initialize variational parameters $\hat{\mu}_k, \hat{R}_k$ for all $k \in \{1, \cdots, K\}$ and $\phi_{t_k}$ for all $t \in \{1, \cdots,T\}$ and $k \in \{1, \cdots, K\}$}
\STATE{}
\WHILE{ELBO has not converged} 
\STATE{}
\FOR{each time step $t \in \{1, \hdots, T\}$}
\FOR{$k \in \{1, \hdots, K\}$}
\STATE{$\phi_{tk} \gets \frac{r_{t(k)}}{\sum_{k=1}^K r_{t_k}}$} \ \ \ \ 
\COMMENT{(where $r_{t_k}$ is as defined in (\ref{rtk1}))}
\ENDFOR
\ENDFOR
\STATE{}
\FOR{$k \in \{1, \hdots, K\}$}
\STATE{$\hat{R}_{k} \gets \left[R_{k0}^{-1}  + M^{-1} \sum_{t=1}^T \phi_{t_k} \right]^{-1}$}
\STATE{$\hat{\mu}_{k} \gets \hat{R}_{k} \left[ R_{k0}^{-1} \mu_{k0} + M^{-1} \sum_{t=1}^T \phi_{t_k} x_t       \right]$}
\FOR{$j \in \{1, \hdots, J\}$}
\STATE{$\hat{\alpha}_{k_j}  \gets \left[ \alpha_{k_j} + \sum_{t=1}^T \phi_{tk} \ d_{t(j)} \right]$ } \text{ (where $d_{t(j)}$ is $1$ if at time $t$ return is in $j'th$ category and $0$ otherwise)}
\ENDFOR
\ENDFOR
\STATE{}
\ENDWHILE
\STATE{}
\RETURN $\phi_{t_k},\hat{R}_{k}, \hat{\mu}_{k}$ and $\hat{\alpha}_{k}$
\end{algorithmic}
\end{algorithm}

\subsection{Determining distribution of $D$ day P\&L and estimating VaR}

In the VI calibration steps described above, it was assumed that both $x_t$ and $d_t$ (category of $D$ day P\&L change starting from time $t$) are known. When we are trying to predict VaR at the current time $t$ while the current market state $x_t$ is known, $d_t$ which describes the future $D$ day P\&L is not known and is indeed what we are trying to predict ($d_t$ will be known only at time $t+D$). The expression for $\phi_{tk}$ in equation (\ref{phitk}) described above requires the knowledge of $d_t$ since the expression for $r_{t_k}$ in (\ref{rtk1}) depends on the $d_t$. Thus for estimating VaR at the current time $t$, the variational parameter corresponding to cluster assignments (corersponding to $\phi_{tk}$) needs to be estimated assuming only $x_t$ is known. Let $\mathcal{D}_t$ be the available data until the current time $t$ from which $D$ day VaR is to be estimated:
\[
\mathcal{D}_t:=\{ \cdots, x_{t-1}, x_t; \cdots, d_{t-D-1}, d_{t-D}\}
\]
We will denote $q_c^{pred}(c_{t(k)}=1)$ as the variational probability of cluster assignment $c_{t(k)}$ based on $\mathcal{D}_t$ (data available only until time $t$). From (\ref{meanfield}) and (\ref{logP}) one obtains the following where $\mathbb{E}_{-c_{t}}$ is expectation with respect to  $\{\mu_1, .., \mu_K \}$ and $\{c_i ; i \neq t \}$
\begin{align}
    q_c^{pred}(c_{t(k)}=1; \phi_t) & \propto \text{exp} \Big( \mathbb{E}_{-c_{t}} \left[ \text{log} \ p(\boldsymbol\mu , \text{\bf{c}}, \mathcal{D}_t) \right]   \Big) \nonumber \\
    & \propto \text{exp} \Big( \mathbb{E}_{-c_{t}} \left[ \text{log} \ p(c_{t(k)}=1,x_t | \pi, \boldsymbol\mu ) \right]  + \text{const} \Big) \nonumber \\
    & \propto \text{exp} \Big( \mathbb{E} \left[ \text{log} \ p(c_{t(k)}=1 | \pi ) \right] + \mathbb{E}_\mu \left[ \text{log} \ p(x_t | c_{t(k)}=1, \mu_k ) \right] + \text{const} \Big) \nonumber \\
    & \propto \text{exp} \Big( \text{log} (\pi_k) - \frac{1}{2} x_t'M^{-1} x_t + x_t'M^{-1} \ \mathbb{E}_\mu( \mu_k) - \frac{1}{2} \mathbb{E}_\mu(\mu_k' M^{-1} \mu_k) \ + \ \text{const} \Big) \nonumber \\
     & \propto \text{exp} \Big( \text{log} (\pi_k) + x_t'M^{-1} \ \mathbb{E}_\mu( \mu_k) - \frac{1}{2} \mathbb{E}_\mu \left[ \text{trace}(M^{-1}\mu_k \mu_k') \right] + \ \text{const} \Big) \nonumber \\
    \label{qpred1}
    & \propto \text{exp} \Big( \text{log} (\pi_k) + x_t'M^{-1} \hat{\mu}_k - \frac{1}{2} \text{trace}(M^{-1} (\hat{\mu}_k \hat{\mu}_k'+\hat{R}_k))
     + \ \text{const} \Big)
\end{align}
where "const" represents terms that are independent of $k$ and $\mu$. One key difference between the above and expression for $q_c^*$ in
(\ref{qcluster}) used in calibration of variational parameters is that expression involving $p(d_{t}|c_{t(k)}=1, \theta_k)$ is not there in (\ref{qpred1}) since $d_{t}$ is not known at time $t$. Since $\sum_k q_c^{pred}(c_{t(k)}) =1$, from (\ref{qpred1}) one observes
\begin{align}
\label{qpred2}
    q_c^{pred}(c_{t(k)}=1) & = \frac{\text{exp} \Big( \text{log} (\pi_k) + x_t'M^{-1} \hat{\mu}_k - \frac{1}{2} \text{trace}(M^{-1} (\hat{\mu}_k \hat{\mu}_k'+\hat{R}_k))
     \Big)}{\sum_{j=1}^K \text{exp} \Big( \text{log} (\pi_j) + x_t'M^{-1} \hat{\mu}_j - \frac{1}{2} \text{trace}(M^{-1} (\hat{\mu}_j \hat{\mu}_j'+\hat{R}_j))\Big)} 
\end{align}
For predicting the conditional probability of $d_{t}$ (category of portfolio value change from $t$ to $t+D$), we will approximate posterior distribution of $\mu$ and $c_t$ by their variational approximation. In particular, the probability of the $J$ possible outcomes for $d_{t}$ are obtained as follows:
\begin{eqnarray}
p(d_{t(j)}=1 \mid \mathcal{D}_t) & =  \
\sum_{k=1}^K  p(d_{t(j)}=1|x_t, c_{t(k)}=1) p ( c_{t(k)}=1 | \mathcal{D}_t)  \nonumber \\
& \approx  \sum_{k=1}^K  q_\theta(d_{t(j)}=1| c_{t(k)}=1) q_c^{pred} (c_{t(k)}=1 )   \nonumber \\
\label{density}
& = \sum_{k=1}^K \left[ \frac{\hat{\alpha}_{k_j}}{\sum_{i=1}^J\hat{\alpha}_{k_i}}  q_c^{pred}(c_{t(k)}=1) \right] 
\end{eqnarray}
where we have used (\ref{catprobvar}) for $q_\theta(d_{t(j)}=1| c_{t(k)}=1)$ and $q_c^{pred} (c_{t(k)}=1 )$ is as defined in (\ref{qpred2}). Below we summarize various steps of the algorithm to estimate distribution of $D$ day P\&L and VaR.

\textbf{Summary of the algorithm to obtain $D$ day P\&L distribution and VaR}

\begin{enumerate}
\item Based on prior $T$ days historical market data used for VaR, obtain empirical distribution $Hist\_Ret_{t}^j$ as defined in (\ref{jcategory}) and $n_{t,j}$ for each of the $J$ categories. 
\item From $T$ most recent observations of $(x_t,d_t)$, obtain the variational parameters $\hat{R}_{k}, \hat{\mu}_{k},\phi_{tk}$ and $\hat{\alpha}_k$ based on CAVI Algorithm \ref{cavialg}.
\item Obtain the cluster probabilities $q_c^{pred} (c_{t(k)}=1 )$ utilizing (\ref{qpred2}) and then obtain the probabilities of $J$ categories of $D$ day P\&L $p(d_{t(j)}=1 | \mathcal{D}_t)$ from (\ref{density}).
\item For $j=\{1,\cdots,J\}$, assign probability to each of the $n_{t,j}$ elements of $Hist\_Ret_{t}^j$ as follows: 
\begin{equation}
\label{histsimprob}
    \text{Probability ($D$ day P\&L is $P\&L_{t,i}$)}=  \sum_{j=1}^J \frac{p(d_{t(j)}=1\mid \mathcal{D}_t)I_{Hist\_Ret_{t}^j}(P\&L_{t,i})}{n_{t,j}} 
\end{equation}
where $I_{Hist\_Ret_{t}^j}(P\&L_{t,i})$ is the indicator function that is $1$ if $P\&L_{t,i} \ \in \ Hist\_Ret_{t}^{j}$ and $0$ otherwise. The above implies that probability of P\&L is non zero only for $P\&L_{t,i} \ \in \ Hist\_Ret_{t}^{j}$ for some $j$ and if $P\&L_{t,i} \ \in \ Hist\_Ret_{t}^{j}$ then its probability is $\frac{p(d_{t(j)}=1\mid \mathcal{D}_t)}{n_{t,j}}$ where $n_{t,j}$ is the number of observations in historical simulation vector $Hist\_Ret_{t}^{j}$.
\item From the probabilities of $T$ discrete outcomes of $Hist\_Ret_{t}$ obtained as in (\ref{histsimprob}), obtain the distribution of $D$ day P\&L and VaR for the desired confidence level.
\end{enumerate}
In the above algorithm there are the same $T$ discrete outcomes for $D$ day P\&L ($P\&L_{t,i}$) as in standard historical simulation for VaR. However, unlike historical simulation where each historical data is assigned equal probability, here the probability for the historical simulated P\&L varies from category to category and is proportional to the likelihood of the $j'th$ category. From (\ref{histsimprob}) one notes that the total probability of all P\&L outcomes in $Hist\_Ret_{t}^j$ is $p(d_{t(j)}=1 \mid \mathcal{D}_t)$ (since there are $n_{t,j}$ elements in $Hist\_Ret_{t}^j$). Thus if category $j$ outcome is more likely ($p(d_{t(j)}=1 \mid \mathcal{D}_t)$ is high), there will be more probability assigned to historical simulated outcomes in category $j$. 

\subsection{Illustrative example of computing VaR for a portfolio of equities and bonds} 

To illustrate the performance of the proposed approach, we consider estimation of one-day and ten-day VaR for a portfolio of equities and bonds. We assume that $50\%$ of the portfolio is invested in S\&P index (equities) and $50\%$ is invested in ten year US Treasuries (ten year US Government bonds). VaR for one-day and ten-days are computed at two confidence levels ($95\%$ and $97.5\%$) using the proposed VI approach and compared to standard Historical Simulation approach as well as Gaussian distribution calibrated to historical simulated portfolio returns. The comparison of VaR approaches was made for the year $2020$ during which we saw both strong sell-offs and rallies in the markets. In February and March of $2020$ there was big sell-off of the risk assets in US due to Covid concerns and then risk assets rallied strongly in the second half due to strong fiscal support from the US government. Due to fluctuations in market conditions, this period provides a good test of how quickly the VaR estimates adapt to changing market conditions.    

For each day, the model is calibrated based on most recent $250$ observations (approximately one year of data) of closing prices of chosen macro indices. For illustrative purposes, only three explanatory variables that capture the momentum effect and recent market volatility are included for cluster identification:

\textit{Market data for one day VaR}
\[
  x_t=\begin{bmatrix}
    \text{Change in ten year US Treasury yield over the most recent day}  \\ \text{VIX index change over the most recent five days} \\ \text{Standard deviation of USD and JPY exchange rate changes over the most recent five days}
    \end{bmatrix}
\]

\textit{Market data for ten day VaR}
\[
  x_t=\begin{bmatrix}
    \text{Change in ten year US Treasury yield over the most recent five days}  \\ \text{U.S. Dollar Index DXY change over the most recent ten days} \\ \text{Standard deviation of USD and JPY exchange rate changes over the most recent five days}
    \end{bmatrix}
\]
Calibration was done after normalizing all inputs and outputs. The first two inputs for both cases and output (portfolio value change over one and ten days) were normalized to their $z$ score based on the previous $250$ observations where $z$ score is defined as (current value-most recent $250$ day average)/standard deviation of variable over the most recent $250$ days. The third input is the difference between the standard deviation of JPY/USD exchange rate over most recent five days and the standard deviation of JPY/USD exchange rate over the most recent $250$ days.

We will consider three clusters ($K=3$) and three categories of outcomes ($J=3$). The three categories of portfolio value changes were based on $D$ day P\&L corresponding to $z$ score changes of $z < -0.8$, $-0.8 \leq z \leq 0.8$ and $z > -0.8$ ($z$ score of P\&L is $D$ day P\&L minus its average over the previous $250$ days divided by the standard deviation of these $250$ P\&L's). These categories correspond to roughly worst $25\%$, middle $50\%$ and top $25\%$ of portfolio value changes.

\begin{figure}[hbt!]
\label{example1D95}
        \includegraphics[width=\textwidth]{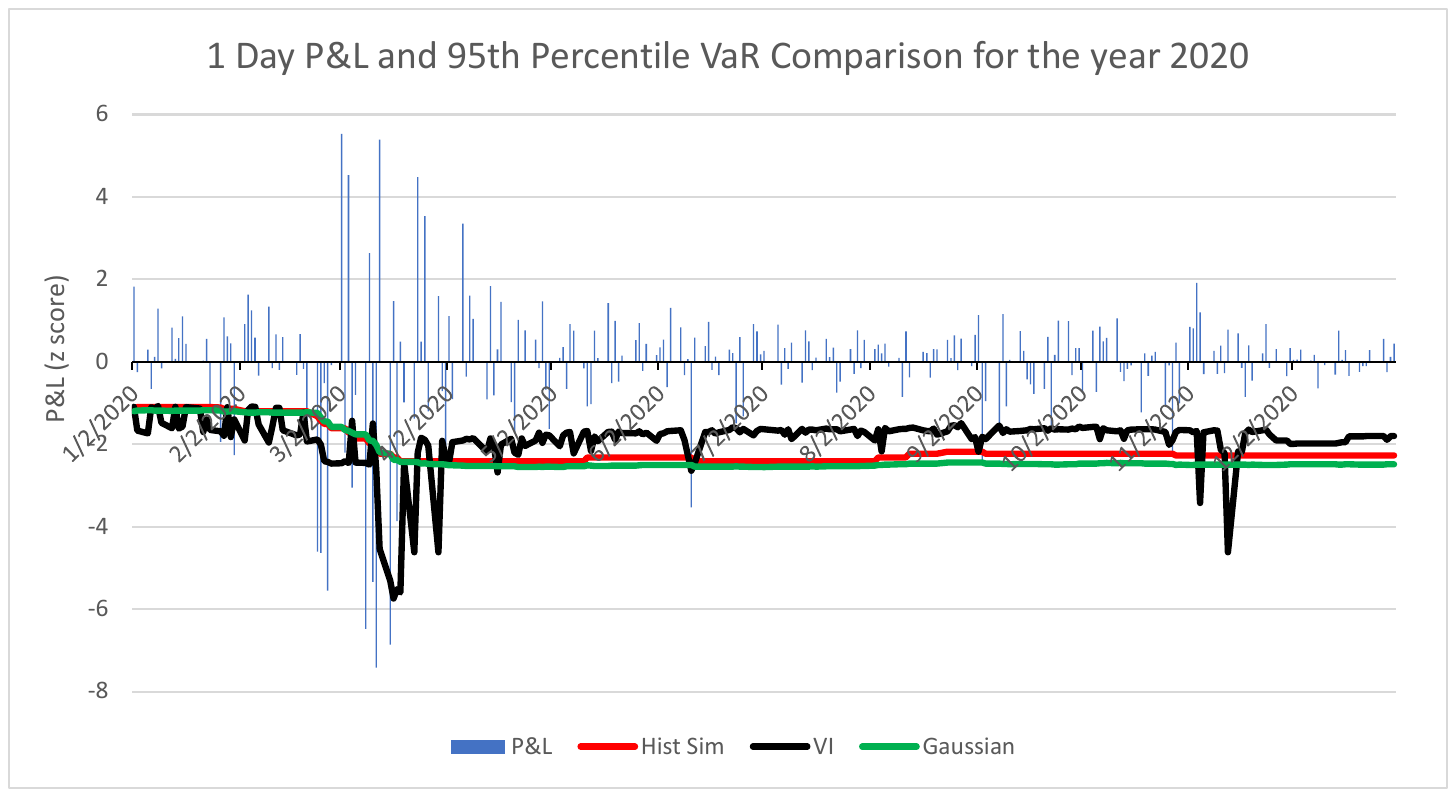}
        \caption{Comparison of $95\%$ One-Day P\&L and VaR for the year $2020$ represented in terms of $z$-scores. VaR derived using the proposed VI approach is compared to that estimated using historical simulation and Gaussian distribution calibrated to historical simulation. VaR estimate in all cases  is based on the same historical data of prior $250$ observations of daily risk factor changes. Both Historical Simulation and Gaussian approaches underestimate VaR in February and March relative to VI. When volatility subsides after May $2020$, VI VaR estimate decreases towards normal levels as it adapts to market conditions while Historical Simulation and Gaussian VaR remain at elevated levels until the volatile historical data is no longer included in the look-back period for VaR estimation.}
        \label{fig:example1D95}
\end{figure}

\begin{figure}[hbt!]
        \includegraphics[width=\textwidth]{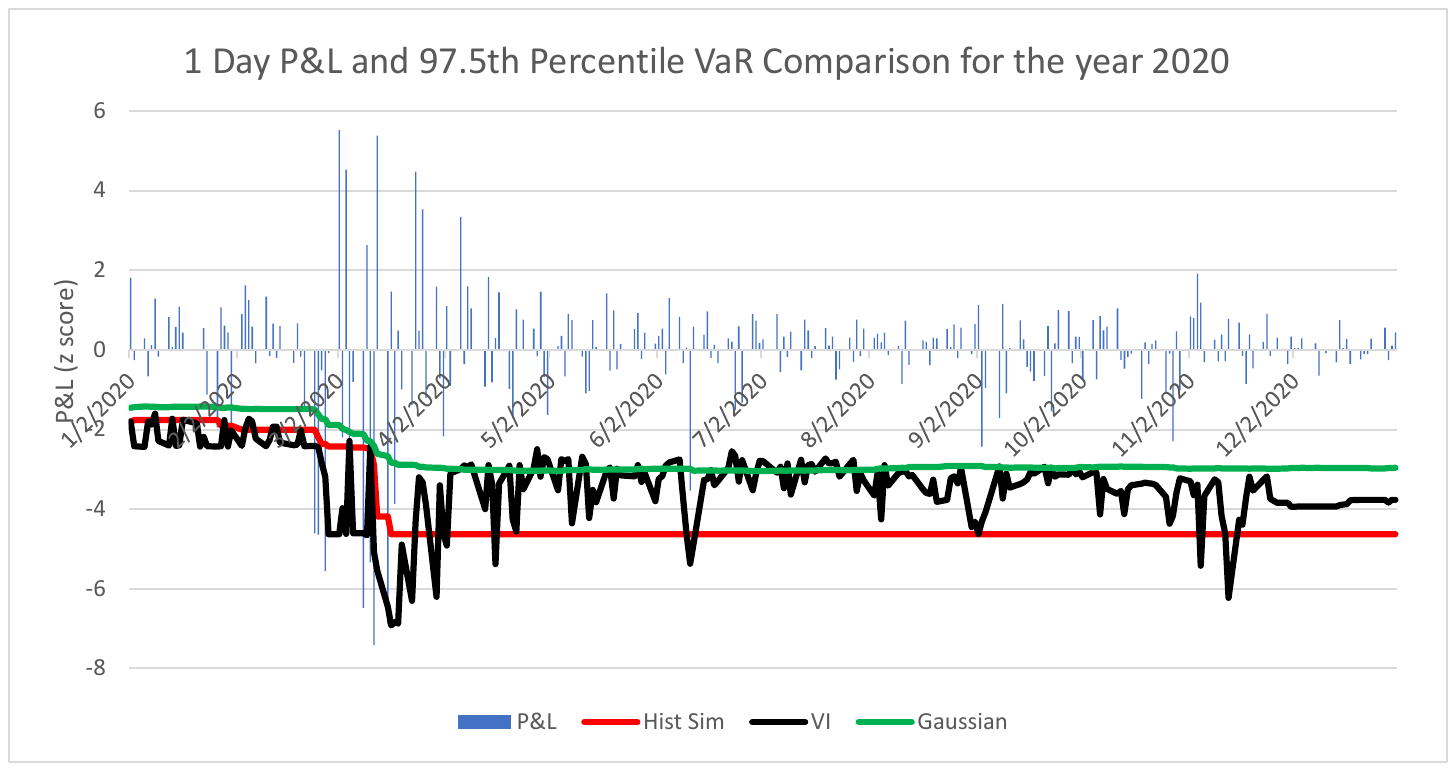}
        \caption{Comparison of $97.5\%$ One-Day P\&L and VaR for the year $2020$ represented in terms of $z$-scores. VaR derived using the proposed VI approach is compared to that estimated using historical simulation and Gaussian distribution calibrated to historical simulation. VaR estimate in all cases  is based on the same historical data of prior $250$ observations of daily risk factor changes. Both Historical Simulation and Gaussian approaches underestimate VaR in February and March relative to VI. Gaussian distribution based VaR is lower than Historical Simulation VaR as Gaussian distribution does not capture the heavy tails of the distribution. When volatility subsides after May $2020$, VI VaR estimate decreases towards normal levels as it adapts to market conditions while Historical Simulation based VaR remains at elevated levels until the volatile historical data is no longer included in the look-back period for VaR estimation. }
        \label{fig:example1D97}
 \end{figure}

Figures \ref{fig:example1D95} and \ref{fig:example1D97} provide comparisons of $1$-Day VaR for $95\%$ and $97.5\%$ confidence levels based on the proposed approaches. VaR estimates are compared to the next day P\&L. Both Historical Simulation and VaR based on Gaussian calibration of Historical Simulation significantly understate VaR in February and March relative to VI when the markets turn volatile with large negative moves in portfolio values. When volatility subsides after May $2020$, VI VaR estimate decreases towards normal levels as it adapts to market conditions while Historical Simulation and Gaussian VaR remain at their respective elevated levels until the volatile historical data is no longer included in VaR estimate. Gaussian parametrized VaR is lower than Historical Simulation VaR for higher confidence level VaR ($97.5\%$ VaR) as Gaussian distribution does not capture the heavy tails of the distribution.


\begin{figure}[hbt!]
        \includegraphics[width=\textwidth]{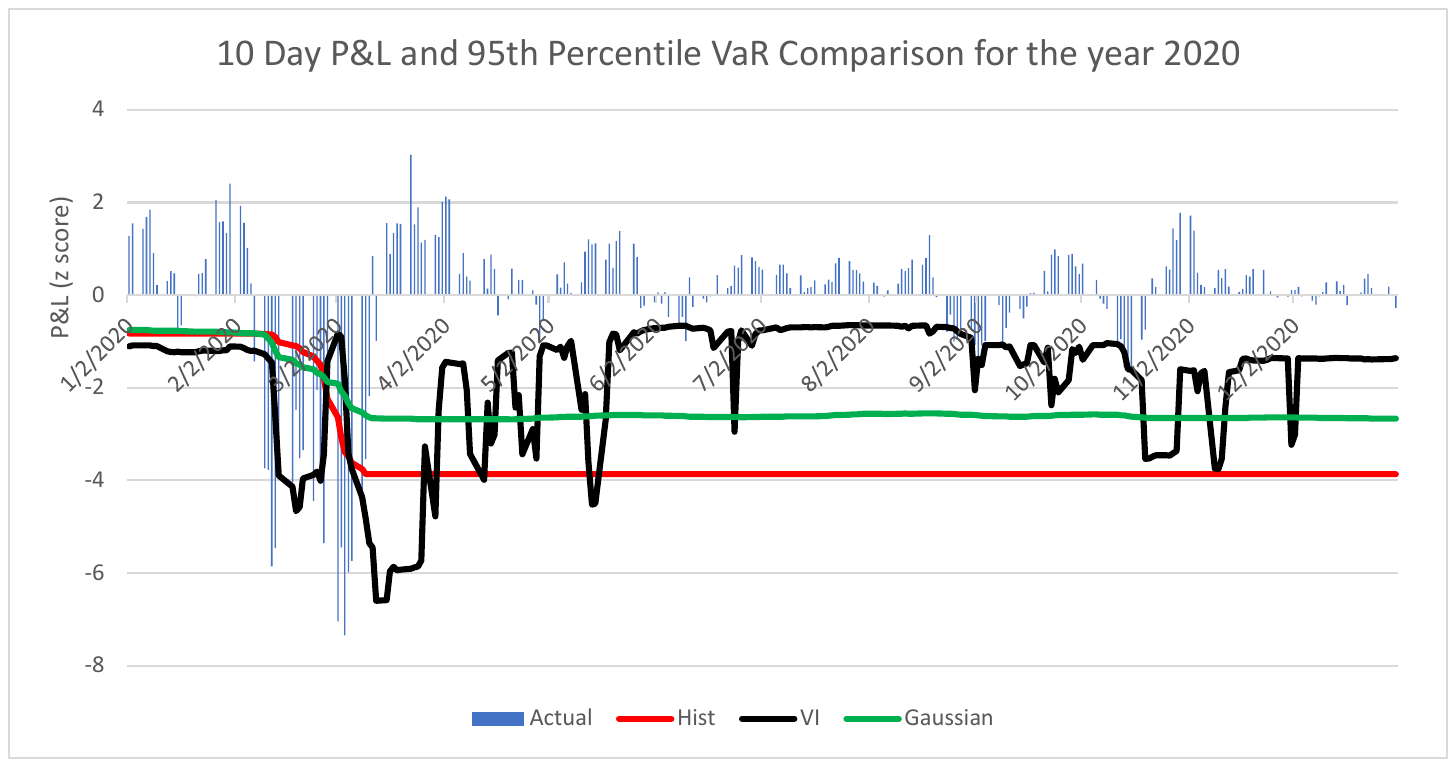}
        \caption{Comparison of $95\%$ Ten-Day P\&L and VaR for the year $2020$ represented in terms of $z$-scores. VaR derived using the proposed VI approach is compared to that estimated using historical simulation and Gaussian distribution calibrated to historical simulation. VaR estimate in all cases  is based on the same historical data of prior $250$ observations of daily risk factor changes. Both Historical Simulation and Gaussian approaches underestimate VaR in February and March relative to VI. Gaussian distribution based VaR is lower than Historical Simulation VaR as Gaussian distribution does not capture the heavy tails of the distribution. When volatility subsides after May $2020$, VI VaR estimate decreases towards normal levels as it adapts to market conditions while Historical Simulation based VaR remains at elevated levels until the volatile historical data is no longer included in the look-back period for VaR estimation.}
  \label{fig:example10D95}
 \end{figure}

\begin{figure}[hbt!]
        \includegraphics[width=\textwidth]{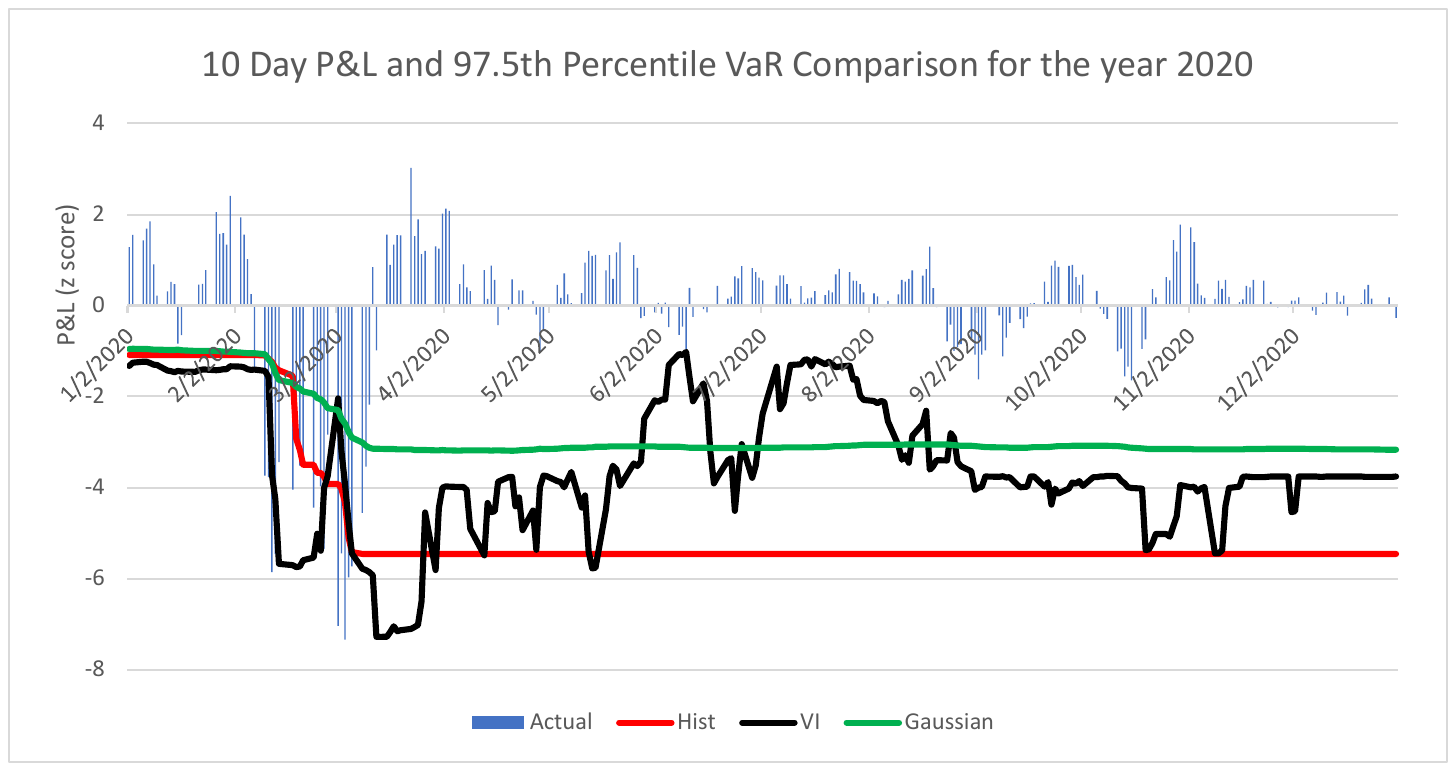}
    \caption{Comparison of $97.5\%$ Ten-Day P\&L and VaR for the year $2020$ represented in terms of $z$-scores. VaR derived using the proposed VI approach is compared to that estimated using historical simulation and Gaussian distribution calibrated to historical simulation. VaR estimate in all cases  is based on the same historical data of prior $250$ observations of daily risk factor changes. Both Historical Simulation and Gaussian approaches underestimate VaR in February and March relative to VI. Gaussian distribution based VaR is lower than Historical Simulation VaR as Gaussian distribution does not capture the heavy tails of the distribution. When volatility subsides after May $2020$, VI VaR estimate decreases towards normal levels as it adapts to market conditions while Historical Simulation based VaR remains at elevated levels until the volatile historical data is no longer included in the look-back period for VaR estimation.}
    \label{fig:example10D975}
 \end{figure}

 Figures \ref{fig:example10D95} and \ref{fig:example10D975} provide a comparison of $10$-Day VaR based on the proposed approaches. Conclusions are similar to those for $1$-Day VaR. As can be seen from these results, VaR based on the proposed approach adapts quickly to chaanging market conditions and VaR estimates are usually more accurate than those computed using Historical Simulation or based on Gaussuain distribution of returns. 
 
 
\clearpage

\section{Stress Scenario Design Incorporating Current Market Conditions}

The underlying framework for stress scenario design is similar to that used for estimating VaR in the previous section but in stress scenario design there are some additional issues to consider:
\begin{itemize}
    \item In stress scenario design we are not only interested in potential loss of portfolio value (as is the case in VaR) but are also interested in obtaining risk factor changes that would result in such a portfolio loss.
    \item In VaR, the time horizon is specified such as one day or ten days. In stress scenarios the horizon of interest is not that precise. In other words future horizon of occurrence (could stress period occur in the next month or the next year) or the duration of stress period from the beginning to end are not precisely specified. 
    \item Sometimes stress scenarios are designed for special market conditions. For example stress scenarios maybe designed for analyzing portfolio risk arising from debt default of country A or a crisis in a particular sector such as real estate or from sudden large increase in rates.
\end{itemize}

The proposed approach utilizes historical data as in the scenario design approach proposed in Nagpal (\cite{nagpal}). The approach in \cite{nagpal} did not explicitly incorporate the current market conditions in the scenario design, while the proposed approach here incorporates them by designing scenarios based on their likelihood of their occurrence in the current market conditions. Thus in the proposed approach, severity of stress scenario and risk factor shocks change with changing market conditions. The main advantages of the proposed approach are:
\begin{itemize}
    \item Stress scenario is designed for current market conditions and thus provides more realistic measures of risk over the near term
    \item Stress scenario is designed specifically for the portfolio of interest (stress scenarios will have different risk factor shifts for different portfolios) 
    \item Adjustable level of severity with objective measures of plausibility of the stress scenario
    \item Incorporates horizon of interest (for example possibility of occurrence in the next month or the next three months) and the duration of the stress period from beginning to end (for example ten days from peak to trough). 
\end{itemize}
To incorporate the features described above, the stress scenario specifications will include three user specified inputs 
\begin{enumerate}
    \item Desired measure of plausibility (what is the likelihood of such a loss which in turn impacts loss severity of the stress scenario)
    \item Horizon of occurrence of the stress event (can we expect such a stress scenario in the next month or the the next six months?)
    \item Maximum duration from the start to end of the stress scenario 
\end{enumerate}
For example we may want to design a scenario for the given portfolio such that \textit{"the probability of loss exceeding the scenario loss over any period of L days or less (for example $15$ business days or less) anytime over the next H days (for example within the next three months) is less than $10\%$"}. In the following we will denote the peak loss in the near future (over the horizon of next $H$ days) by $\text{Loss}^t_{L,H}$:

\begin{equation}
\label{Lthdef}
    \text{Loss}^t_{L,H}:=\text{Maximum loss over any period of length $L$ days or less between time $t$ and $t+H$}
\end{equation}

For example, $\text{Loss}^t_{15,90}$ represents maximum loss in portfolio value over any period of $15$ days or less between day $t$ and day $t+90$. $\text{Loss}^t_{L,H}$ is known only at time $t+H$ and in stress scenario design at time $t$, one of the key objectives is to have good estimate of $\text{Loss}^t_{L,H}$ at time $t$. The proposed approach in this section provides a) distribution of $\text{Loss}^t_{L,H}$ based on market conditions at time $t$, and b) expected change in market risk factors that would produce a portfolio loss for the chosen target scenario loss from the distribution of $\text{Loss}^t_{L,H}$. 

In the following sections we will propose algorithms for two types of scenario designs that were also considered in \cite{nagpal} with the main difference being that here we want to also incorporate the current market conditions in the scenario design. The first type of scenario design is for near term stress scenario design without any constraints where we consider the problem "what is the maximum loss we could expect to see over any $L$ day period in the next $H$ days?" The second type of stress scenario design considered in this paper is for particular type of market conditions (for example, what is the maximum loss we could expect to see over any $L$ day period in the next $H$ days in an environment of increasing rates). These two scenario design objectives are described below:

\begin{itemize}
    \item \textit{Unconstrained scenario design for the current market condition}: For a specified confidence level $p\%$, determine the potential portfolio loss $x$ such that the probability of $\text{Loss}^t_{L,H}$ exceeding $x$ is $p\%$. Obtain the expected value of all risk factor shocks for the portfolio loss of $x$ under the current market conditions.  
    \item \textit{Constrained scenario design for the current market condition}: The same objective as above under some constraints on market changes for which the scenario has to be designed (e.g., interest rates increase and/or oil prices fall). As an example the scenario design for increasing interest rates market conditions could be - for a specified confidence level $p\%$, determine the potential portfolio loss $x$ such that in a scenario with increasing interest rates, the probability of $\text{Loss}^t_{L,H}$ exceeding $x$ is $p\%$. Obtain the expected value of all risk factor shocks for the portfolio loss of $x$ when the interest rates increase.
\end{itemize}

\subsection{Underlying Model for Data Generation}

Since the objective in stress scenario design is to estimate potential portfolio losses, the goal will be to develop a model for the distribution of $\text{Loss}^t_{L,H}$. While in the previous section for VaR the model linked clusters based on market conditions to portfolio returns, here clusters capture similar outcomes for $\text{Loss}^t_{L,H}$. The additional complication in scenario design arises from the fact that apart from estimation of loss severity ($\text{Loss}^t_{L,H}$) for a desired confidence level, we have to also estimate plausible market risk factor changes that would produce the specified portfolio loss. Thus apart from the distribution of $\text{Loss}^t_{L,H}$, we also need the conditional distribution of risk factor changes for the portfolio loss $\text{Loss}^t_{L,H}$. In this section we will describe the assumed underlying model for $\text{Loss}^t_{L,H}$ and changes in market risk factors (such as change in interest rates, equity prices, volatilities etc.) that are used in determining the portfolio value.

Since the approach is based on historical data, to calibrate the model we need to first obtain $\text{Loss}^t_{L,H}$ for for each date $t$ in historical data used for scenario design. For the historical data, $\text{Loss}^t_{L,H}$ can be obtained from the daily portfolio value change based on daily risk factor changes as $\text{Loss}^t_{L,H}$ is the the maximum loss in portfolio value over any horizon of $L$ days or less over the time period $t$ to $t+H$. This is just obtained by sorting as follows where $\text{P\&L}_{t_1,t_2}$ represents the portfolio value change based on change in market risk factors from $t_1$ to $t_2$:
\begin{align}
    \text{Peak\_Loss}_t & =\text{Minimum}\{\text{P\&L}_{t,t+1}, \ \text{P\&L}_{t,t+2}, \, \hdots,  \text{P\&L}_{t,min(t+H,t+L)} \}  \nonumber \\
    \label{Lcomp}
    \text{Loss}^t_{L,H} & =\text{Minimum}\{\text{Peak\_Loss}_t, \ \text{Peak\_Loss}_{t+1}, \ \hdots, \text{Peak\_Loss}_{t+H-1} \}
\end{align}

In the above $\text{Peak\_Loss}_t$ is the maximum loss for all periods with start date of $t$ and end date of no later than $t+L$. $\text{Loss}^t_{L,H}$ is the maximum loss over all possible start dates within the next $H$ days of $t$. Thus utilizing historical market data, from (\ref{Lcomp}) we can obtain $\text{Loss}^t_{L,H}$ for all days in historical data that would be used for scenario design. Since in scenario design we are also interested in estimating changes in market risk factors corresponding to the portfolio loss, from the historical data we will also extract all the relevant risk factor changes over the period where portfolio loss would have been $\text{Loss}^t_{L,H}$. We will use the following notation for risk factor changes over the period corresponding to portfolio loss $\text{Loss}^t_{L,H}$:

\begin{equation}
\label{Rthdef}
    \text{RF}^i_{t,L}:=\text{Shift in risk factor $i$ between the start and end dates of }\text{Loss}^t_{L,H}
\end{equation}

Our data generation framework is similar to that considered in Figure 1 with a few differences. Each day it is assumed that the market is in one of $K$ possible clusters (which are hidden/latent variables) which depend on market data $x_t \in \mathbb{R}^n$ which is known at time $t$. Elements of $x_t$ (such as rate, equity and volatility trends) have been chosen to provide good differentiation of $\text{Loss}^t_{L,H}$ and the accompanying market conditions. From each of the $K$ clusters, there are $J$ possible categories of outcomes. While for VaR the $J$ categories of outcomes were defined only in terms of portfolio loss magnitude, the $J$ categories of outcomes in scenario design may also include key nature of key risk factor changes during the stress period as our goal is to not only estimate potential loss in the stress scenario but also estimate the risk factor changes that would result in such a loss. For example if significant portfolio losses can occur in both increasing as well as decreasing interest rate environments, it would be important to distinguish them so that at any time one can determine whether the stress scenario with increasing or decreasing interest rates is more likely. Continuing this example, if interest rate is an important risk factor for the portfolio and significant losses in portfolio can occur in both increasing and as decreasing interest rate environments, the $J$ category outcomes could based on the six combinations of loss severity and interest rate changes over the period of maximum loss:

\[
\text{Loss category defined in terms of $\{Loss \ severity \ , \ Interest \ rate \ change\}$}
\]
\[
\text{where } \ \{Loss \ severity \} \ = \{ High, \ Medium, \ Low \} \; \; \text{and } \{ Interest \ rate \ change \} \ = \{Increase, \ Decrease\}
\]
Here the six categories ($J=6$) of possible outcomes are 1) high loss severity with rate decrease, 2) medium loss severity with rate decrease, 3) low loss severity with rate decrease, 4) high loss severity with rate increase, 5) medium loss severity with rate increase, and 6) low loss severity with rate increase. The thresholds for loss severity and the key risk factor changes (interest rate in the example above) for the $J$ categories are user specified. Let $i=key$ be the key risk factor in classification of the stress loss category. Let $Loss_{min}^j$ and $Loss_{max}^j$ be the thresholds for $\text{Loss}^t_{L,H}$ for the $j'th$ category and $\text{RF}^{key}_{j, min}$ and $\text{RF}^{key}_{j, max}$ be the thresholds for the key risk factor change over the period $\text{Loss}^t_{L,H}$. In other words for a historical period date $t$, the near term peak loss is in the $j'th$ category if $\text{Loss}^t_{L,H}$ is within the required interval and the key risk factor changes over the period of $\text{Loss}^t_{L,H}$ are also in the appropriate interval:
\[
\text{Loss}^t_{L,H} \text{ in category $j$ if }  \; \Biggl\{ \begin{matrix} Loss_{min}^j \leq \text{Loss}^t_{L,H} \leq Loss_{max}^j \\ \text{ and }  \\ \text{RF}^{key}_{j, min} \leq \text{RF}^{key}_{t,L} \leq \text{RF}^{key}_{j, max} \end{matrix} 
\]
While the above category classification includes both portfolio loss severity as well the nature of market changes for risk factors, it is not necessary to include risk factor changes in category classification. The thresholds used for classification of loss severity and key risk factor changes (such as interest rates above) are user specified and chosen based on historical data so that there are sufficient observations in each of the $j$ categories. The thresholds for the $J$ categories have to be chosen so that any possible outcome will fall into only one of the $J$ categories.  

Utilizing the historically observed risk factor shifts and $\text{Loss}^t_{L,H}$ determined based on those, one can obtain the empirical distribution of $\text{Loss}^t_{L,H}$ for each of the $j$ categories. While other distributions can be considered, we will assume Gaussian distribution for $\text{Loss}^t_{L,H}$ represented by $\mathcal{N}(\bar{L}_j,\sigma_j^2)$ where the parameters $\bar{L}_j$ and $\sigma_j$ representing mean and standard deviation of $\text{Loss}^t_{L,H}$ are obtained from historical observations of $\text{Loss}^t_{L,H}$ in category $j$. To obtain the expected risk factor shocks consistent with the stress scenario loss, one also needs to know the joint distribution of portfolio losses and risk factor changes. For each of the $J$ categories we will assume that the joint distribution of each risk factor and the portfolio loss is a bivariate normal distribution calibrated to historical data of $\text{Loss}^t_{L,H}$ and $\text{RF}^i_{t,L}$ for that particular category. Other distributions can also be considered but assuming bivariate distribution allows for easy estimation of expected risk factor shifts for a specified scenario loss. Let $Q_{L,i}^j$ be the bivariate normal distribution for $\text{Loss}^t_{L,H}$ and $\text{RF}^i_{t,L}$ in category $j$. Thus if there are $N_{RF}$ number of risk factors and $J$ categories, then there are $N_{RF} \times J$ such bivariate distributions. 

\clearpage

Based on the above assumptions, $\text{Loss}^t_{L,H}$ (the peak loss over any period of $L$ days or less over the horizon of the next $H$ days) and the corresponding risk factor shifts $\text{RF}^i_{t,L}$ are generated as follows:
\begin{center}
\textbf{The data generating process for near term peak loss $\text{Loss}^t_{L,H}$ and corresponding risk factor changes $\text{RF}^i_{t,L}$}
\end{center}

\begin{center}
\fbox{%
\parbox{0.8\linewidth}{%
        \begin{itemize}
            \item $J$ is the number of non overlapping categories of outcomes. The $J$ category of loss outcomes are based on peak near term loss ($\text{Loss}^t_{L,H}$) and if needed key risk factor changes in the stress period.
            \item Within each category $j$, stress loss $\text{Loss}^t_{L,H}$ has normal distribution $\mathcal{N}(\bar{L}_j,\sigma_j^2)$. The mean $\bar{L}_j$ and standard deviation $\sigma_j$ are based on historical data of $\text{Loss}^t_{L,H}$ in category $j$.
            \item Within each category $j$, stress loss $\text{Loss}^t_{L,H}$ and each risk factor change have bivariate normal distribution. The bivariate normal distribution for $j'th$ category and $i'th$ risk factor is $Q_{L,i}^j$. The bivariate distribution is calibrated to historical data of $\text{Loss}^t_{L,H}$ and $\text{RF}^i_{t,L}$ for each category $j$ and each risk factor $i$.
            \item $K$ is the number of clusters of input space $x_t$ (where $x_t \in \mathbb{R}^{n}$ and is known at time $t$). 
            \item The fraction of times $x_t$ is in cluster $K$ is $\pi_k$. Thus the sum of $\pi_k$ over all $k$ is one.
            \item For cluster $k$, the mean of $x_t$ is $\mu_k$, which is normally distributed with mean $\mu_{k0}$ and variance $R_{k0}$. In other words, for cluster $k$, $\mu_k \sim \mathcal{N}(\mu_{k0},R_{k0})$. 
            \item If $x_t$ is in cluster $k$ then $x_t$ is normally distributed with  mean $\mu_{k}$ and variance $M$. In other words if $c_{t(k)}=1$ then $x_t \sim \mathcal{N}(\mu_{k},M)$. Note that while each cluster has a different mean $\mu_k$, we are assuming the same variance $M$ for $x_t$ in each cluster.
            \item For cluster $k$, the proportions vector $\theta_k$ describes proportion of times $D$ day portfolio return is in each of the $J$ categories (note $\sum_{j=1}^J \theta_{k_j}=1$). The prior for proportions $\theta_k$ for cluster $k$ is $J$-Dirichlet distribution with parameters $\alpha_k$. 
        \end{itemize}
        \textbf{Generative process for $\text{Loss}^t_{L,H}$ and the corersponding risk factor changes}
        \begin{enumerate}
            \item For $\; k \in \left[ 1, \hdots, K \right]$, draw $\mu_k \ \sim \mathcal{N}(\mu_{k0},R_{k0})$
            \item For $k \in \left[ 1, \hdots, K \right]$, draw proportions $\theta_k \sim Dirichlet(\alpha_k)$. (note $\sum_{1}^J \theta_{k_j}=1$)
            \item For $t \in \left[ 1, \hdots T \right]$   
            \begin{enumerate}
                \item Draw latent variable $c_{t(k)} \ \sim \ \text{Cat} (\pi_k)$ which describes the cluster assignment for time $t$. $c_{t(k)}$ is an indicator variable which is 1 for only one of its $K$ elements and zero for all other $K-1$ elements.
                \item If the cluster assignment in previous step is $k$, draw input $x_t  \ \sim \mathcal{N}(\mu_{k},M)$
                \item If the cluster assignment is $k$, draw stress loss category change category assignment $d_t \ | \theta_k \sim \text{Cat}(\theta_k)$. ($d_{t(j)}$ is $1$ for only one of the $J$ elements of $d_t$).
                \item If the portfolio change drawn in the previous step is in category $j$, draw $\text{Loss}^t_{L,H}$ and from distribution $\mathcal{N}(\bar{L}_j,\sigma_j^2)$.
                \item If the portfolio change is in category $j$, for each risk factor $i$ obtain $\text{RF}^i_{t,L}$ from bivariate distribution $Q_{L,i}^j$ conditioned on $\text{Loss}^t_{L,H}$ drawn in the previous step. 
            \end{enumerate}
        \end{enumerate}
    }
}
\end{center}

\clearpage

\subsection{Variational Inference Calibration for Stress Scenario Design}

There are differences in data generation processes for VaR and stress scenario but they do not impact estimation of the VI parameters. For example in VaR, the $D$-day P\&L was based on discrete outcomes from historical outcomes while in the above $\text{Loss}^t_{L,H}$ is drawn from $\mathcal{N}(\bar{L}_j,\sigma_j^2)$. While the distribution of P\&L is different, in both cases the distribution of P\&L or loss within each category $j$ is assumed to be known (since it is derived from historical experiences). From historical data of $x_t$ and $d_t$ (the category of stress loss outcome), the goal as in the case of VaR is to obtain the probability of cluster assignments for each $t$, the mean and variance of $x_t$ in each cluster, the probabilities of $J$ category outcomes from each cluster. From the historical data of $x_t$ and $d_t$ used for scenario design, the VI parameters $\phi_{t_k},\hat{R}_{k}, \hat{\mu}_{k}$ and $\hat{\alpha}_{k}$ can be obtained as as described in the CAVI Algorithm \ref{cavialg}.

\subsection{Stress Scenario Design : determining target scenario loss and corresponding risk factor shifts}

In designing the stress scenario, we have to determine what the expected portfolio loss in the stress scenario as well as the risk factor shifts likely to produce that scenario loss. Once all the VI parameters have been obtained and probabilities computed, the distribution of $\text{Loss}^t_{L,H}$ can be obtained as follows:
\begin{align}
    p(\text{Loss}^t_{L,H}=L) &  =  \sum_{j=1}^J  p(\text{Loss}^t_{L,H}=L \mid d_{t(j)}=1 ) p(d_{t(j)}=1) \nonumber \\
    \label{stressLdistrib}
    & = \sum_{j=1}^J \  p \left( \text{Loss}^t_{L,H}=L \mid  d_{t(j)}=1 \right) 
\left[ \sum_{k=1}^K p \left( d_{t(j)}=1 \mid c_{t(k)}=1 \right) p\left(c_{t(k)}=1 \right)  \right]
\end{align}
From the above, one can obtain target scenario loss $L^*$ for the desired confidence level. If we want to design a stress scenario such that near term peak loss should be lower than scenario loss with probability $p\%$, then we would determine $L^*$ such that $Prob(\text{Loss}^t_{L,H} \leq L^*)=p$. To determine scenario shifts for the desired scenario loss of $L^*$ we will use the following well known conditional estimate for Gaussian random variables:
\begin{prop}
Let $X$ and $Y$ be Gaussian random variables with the following mean and variances:
\[
\mathbb{E}\{X\}=\bar{X}, \; \mathbb{E}\{Y\}=\bar{Y}, \; \mathbb{E}\{(X-\bar{X})^2\}=\Sigma_{xx}, \; \mathbb{E}\{(X-\bar{X})(Y-\bar{Y})\}=\Sigma_{xy}
\]
Then the conditional estimate of $Y$ given $X$ is
\begin{equation}
    \label{condest}
   \mathbb{E}\{Y \mid X\}=\bar{Y} +\frac{\Sigma_{xy}}{\Sigma_{xx}}(X-\bar{X})
\end{equation}
\end{prop}
The above result can be used to obtain risk factor shifts for desired loss level $L^*$ and given category of loss. In particular $\mathbb{E}\{\text{RF}^i_{t,L} \mid d_{t(j)}=1, \text{Loss}^t_{L,H}=L^*\}$, the conditional estimate with bivariate Gaussian distribution $Q_{L,i}^j$ can be computed using (\ref{condest}) . With this conditional estimate, risk factor shifts for all risk factors can be computed as follows:
\begin{equation}
\label{RFestimate}
    \mathbb{E}\{\text{RF}^i_{t,L} \mid \text{Loss}^t_{L,H}=L^* \}=\sum_{j=1}^J p(d_{t(j)}=1) \  \mathbb{E}\{\text{RF}^i_{t,L} \mid d_{t(j)}=1, \text{Loss}^t_{L,H}=L^*\}
\end{equation}

where the expectation $\mathbb{E}\{\text{RF}^i_{t,L} \mid d_{t(j)}=1, \text{Loss}^t_{L,H}=L^*\}$ is computed using (\ref{condest}) where the mean and variance are those the bivariate normal distribution $Q_{L,i}^j$.

\textbf{Summary of the algorithm to obtain target stress loss and risk factor shifts}

Below we summarize the steps discussed above for scenario design for confidence level $p^*$ (scenario to be designed for loss severity $L^*$ so that $Prob(\text{Loss}^t_{L,H} \leq L^*)=p^*$:

\begin{enumerate} 
\item {\em Obtaining distributions for each of the categories based on historical data}: For all the days $t$ in historical data, obtain $\text{Loss}^t_{L,H}$ from historical data as described in (\ref{Lcomp}). For the time interval corresponding to loss $\text{Loss}^t_{L,H}$, obtain risk factor shifts $\text{RF}^i_{t,L}$ for each risk factor $i$. Based on $\text{Loss}^t_{L,H}$ and $\text{RF}^i_{t,L}$, obtain $d_t$ (the category classification) for all $t$. Obtain parameters of normal distribution $\mathcal{N}(\bar{L}_j,\sigma_j^2)$ (distribution of $\text{Loss}^t_{L,H}$ in $j'th$ category) in each of the $J$ categories. For each risk factor $i$ and each category $j$ obtain the bivariate normal distribution $Q_{L,i}^j$ from observations in $j'th$ category of $\text{Loss}^t_{L,H}$ and $\text{RF}^i_{t,L}$.
 \item {\em Determine variational parameters from historical data}: From the historical data of $(x_t,d_t)$ used for scenario design, obtain the variational parameters $\hat{R}_{k}, \hat{\mu}_{k},\phi_{tk}$ and $\hat{\alpha}_k$ using the CAVI Algorithm \ref{cavialg}.
\item {\em Estimate the cluster and category probabilities for the most recent date}: For the current time step, obtain the cluster probabilities $q_c^{pred} (c_{t(k)}=1 )$ utilizing (\ref{qpred2}) and then obtain the probabilities of $J$ categories of stress loss outcomes $p(d_{t(j)}=1 | \mathcal{D}_t)$ from (\ref{density}).
\item {\em Obtain the distribution of stress loss}: Obtain the distribution of $\text{Loss}^t_{L,H}$: 
\[
p(\text{Loss}^t_{L,H}=L)  =  \sum_{j=1}^J p(d_{t(j)}=1) \ p(\text{Loss}^t_{L,H}=L \mid d_{t(j)}=1 )
\]
where from (\ref{density}) and specified distribution of $\text{Loss}^t_{L,H}$ in category $j$ one notes
\[
p(d_{t(j)}=1)=\sum_{k=1}^K \left[  \frac{\hat{\alpha}_{k_j}}{\sum_{i=1}^J\hat{\alpha}_{k_i}}  q_c^{pred}(c_{t(k)}=1) \right] \;, \; \text{and}
\]
\[
p(\text{Loss}^t_{L,H}=L \mid d_{t(j)}=1 ) =\frac{1}{\sigma_j \sqrt{2 \pi}} \text{exp}^{-\frac{1}{2}(\frac{L-\bar{L}_j}{\sigma_j})^2}
\]

\item {\em Obtain the target stress loss for the desired confidence level}: From the distribution of $\text{Loss}^t_{L,H}$, obtain the target scenario loss $L^*$ for the desired confidence level $p^*$. For example if we want a scenario so that with $95\%$ probability $\text{Loss}^t_{L,H}$ is below the stress loss, then $L^*$ is obtained so that probability of $\text{Loss}^t_{L,H}$ being lower than $L^*$ is $95\%$.
\item {\em Obtain risk factor shifts for the target scenario loss}: For each risk factor $i$, obtain the risk factor shift using (\ref{RFestimate}) where the expectation $\mathbb{E}\{\text{RF}^i_{t,L} \mid d_{t(j)}=1, \text{Loss}^t_{L,H}=L^*\}$ is with respect to bivariate Gaussian distribution $Q_{L,i}^j$. This expectation can be computed using  equation (\ref{condest}).
\end{enumerate}

\subsection{Scenario Design for Specific Market Conditions}

In many situations we are interested in potential losses in portfolio value in a particular market environment. For example one may wish to create a scenario with large increase in crude oil prices if that was a particular concern, or create a scenario with significant depreciation of some currency. We will extend the above algorithm to such constrained scenario design as in \cite{nagpal}. The approach will still rely on historical data where only that subset of historical data is used during which market conditions were consistent with the desired market conditions. Thus if for example we are interested in scenario design with increasing crude oil prices, as in \cite{nagpal} our historical data for calibration will be limited to only those combinations of historical start and end dates during which crude oil prices rose. 

To illustrate the approach we will assume that the scenario of interest is described in terms of inequality constraint of $key'th$ risk factor shift as
\begin{equation}
    \label{scen_shift_def}
    \text{RF}^{key} \geq c^* \; \; (\text{The inequality constraint can also be } \text{RF}^{key} \leq c^* )
\end{equation}
For example if we are interested in stress scenario with increasing crude oil prices, the constraint may be that crude oil prices increase at least $5\%$ in the designed stress scenario. There is no restriction that scenario of interest be defined in terms of only one risk factor shift - all we need is an ability to identify historical periods which satisfy the desired scenario attributes. With this constraint in mind, we will define peak near term loss as the peak loss where the start and end times are such that the risk factor constraints are satisfied: 
\begin{align}
\text{Eligible\_end\_dates}_t & = \{ t_1 : t_1 \leq t+L, \ t_1 \leq t+H \text{ and risk factor shifts from $t$ to $t_1$ satisfy constraints (\ref{scen_shift_def})} \} \nonumber \\
 \text{Peak\_Loss}_t & =\text{Minimum}\{\text{P\&L}_{t,t_1} \ : \ t_1 \in \ \text{Eligible\_end\_dates}_t \} \nonumber \\
\text{Loss}^t_{L,H} & =\text{Minimum}\{\text{Peak\_Loss}_t, \ \text{Peak\_Loss}_{t+1}, \ \hdots, \text{Peak\_Loss}_{t+H-1} \} \nonumber \\
\text{RF}^i_{t,L} & =\text{Shift in risk factor $i$ between the start and end dates of }\text{Loss}^t_{L,H} \nonumber
\end{align}
Here $\text{Eligible\_end\_dates}_t$ is the set of end dates so that the key risk factor shift from $t$ to that date satisfies the desired scenario constraints. With $\text{Loss}^t_{L,H}$ and $\text{RF}^i_{t,L}$ obtained as above, we can proceed as in the previous section to design the stress scenario for the specified confidence level.  

\subsection{Illustrative example of stress scenario design} 

To illustrate the performance of the proposed approach, we consider unconstrained scenario design for a portfolio of equities and bonds. We assume that $50\%$ of the portfolio is invested in S\&P index (equities) and $50\%$ is invested in ten year US Treasuries (ten year US Government bonds). The stress scenarios are to be designed for peak loss that could occur in a duration of fifteen business days or less ($L=15$ days in the notation above) over the horizon of the next forty five business days ($H=45$ days in the notation above). The scenarios are to be designed for two levels of loss severity of $75\%$ and $95\%$. This implies that scenario designs should be such that with probability of $75\%$ and $95\%$, $\text{Loss}^t_{15,45}$ is lower than scenario loss.      

For illustrative purposes, only three explanatory variables that capture the momentum effect and recent market volatility are included for cluster identification:

\textit{Market data used for stress scenario design}
\[
  x_t=\begin{bmatrix}
    \text{Most recent five day average of US Dollar DXY index relative to its previous one year average}  \\ \text{Most recent five day average of VIX index relative to its previous one year average} \\ \text{Most recent ten day average value of portfolio value relative to its last year average}
    \end{bmatrix}
\]

All the variables are normalized to their $z$ score ($z$ score of a variable is its value minus its previous year average divided by its one year standard deviation). We will assume that for $x_t$ defined above, there are four clusters ($K=4$). For the chosen portfolio, in recent years there have been periods of significant portfolio loss in both increasing as well as decreasing interest rate environments. For this reason we would like to be able to distinguish stress scenarios with increasing rates from those in which interest rates decrease. To incorporate direction of this key risk factor change in scenario design, will define stress categories in terms of both portfolio loss severity as well as change in interest rates in the stress period. We will assume there are eight categories of stressed outcomes ($J=8$ in the notation above) - five with different levels of loss severities where in each case interest rates decrease during the stress period and three categories with different levels of loss severities where interest rates increase. 

\begin{figure}[hbt!]
        \includegraphics[width=\textwidth]{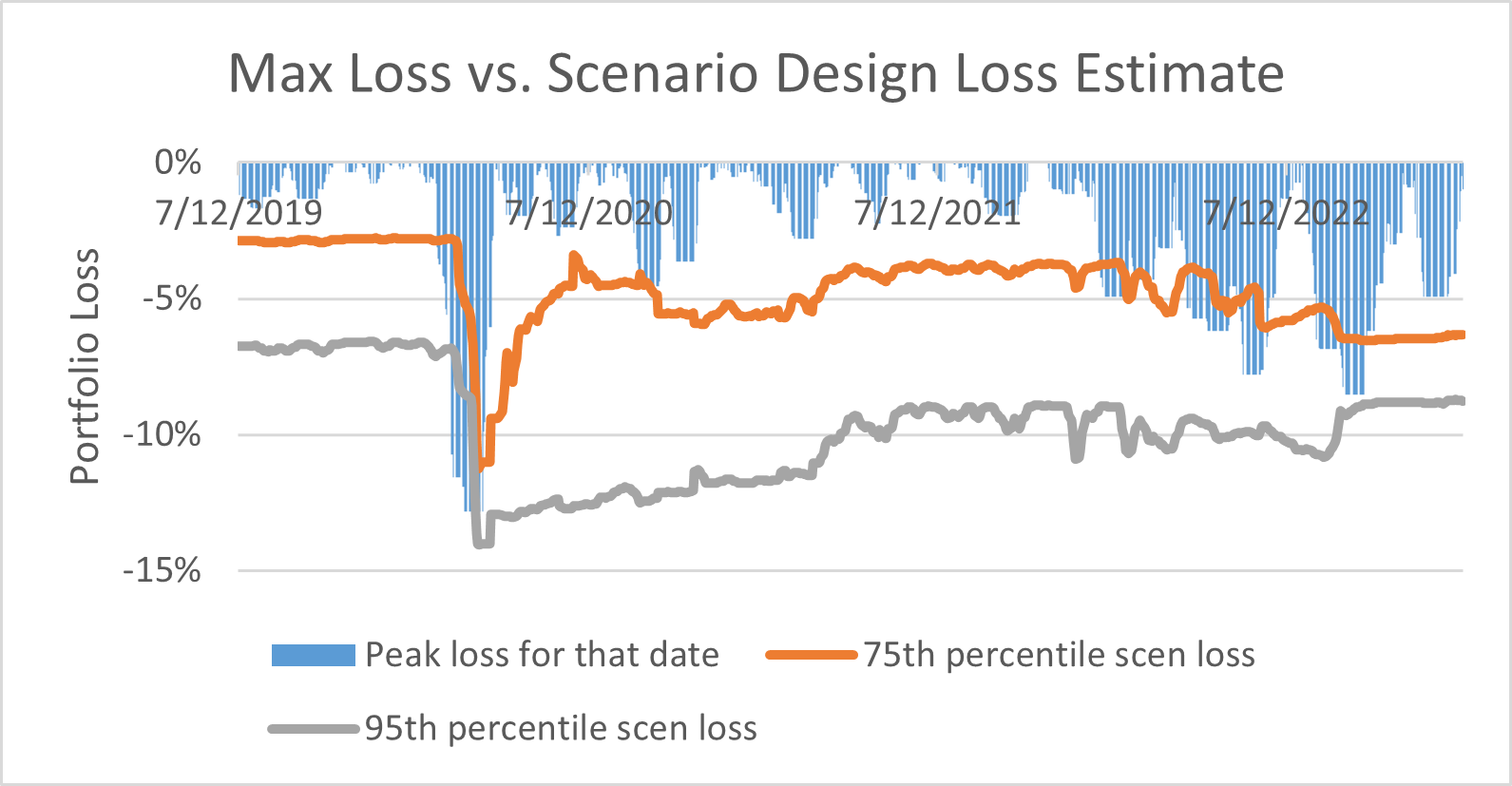}
        \caption{Comparison of peak loss for each day vs. target scenario loss for $75\%$ and $95\%$ confidence level from mid $2017$ to end of year $2022$. Peak loss for  date is the maximum loss in any period of fifteen days or less that includes that date. The target scenario losses are such that with probability of $75\%$ and $95\%$ respectively, $\text{Loss}^t_{15,45}$ (peak loss within $15$ days over the next $45$ days) is expected to be lower than scenario loss. Note that the target scenario losses quickly increase in spring of $2020$ during the period of heightened Covid concern and then decline to lower levels as markets stabilize. The target scenario losses for the less severe scenario ($75\%$ confidence level) again increase in the year $2022$ during the period of volatility.}
  \label{fig:losstarget}
 \end{figure}

For each day the model is calibrated based on most recent $1000$ daily observations (approximately four years of data) preceding that day. Based on the algorithm described in Section 4.3, for each day a stress scenario is designed where we obtain the target loss level for the stress scenario and the corresponding shifts in US ten year Treasury yields and S\&P index (the two risk factors for the portfolio). Figure \ref{fig:losstarget} shows peak losses in a period of fifteen days or less that includes that date vs. the loss severity of scenario designs for that date. While the target scenario loss in January $2020$ not predict stress losses February-March $2020$, the projected scenario losses increase rapidly in February-March $2020$ as markets become volatile and subside as markets stabilize in the summer of $2020$. 
 
\begin{figure}[hbt!]
        \includegraphics[width=\textwidth]{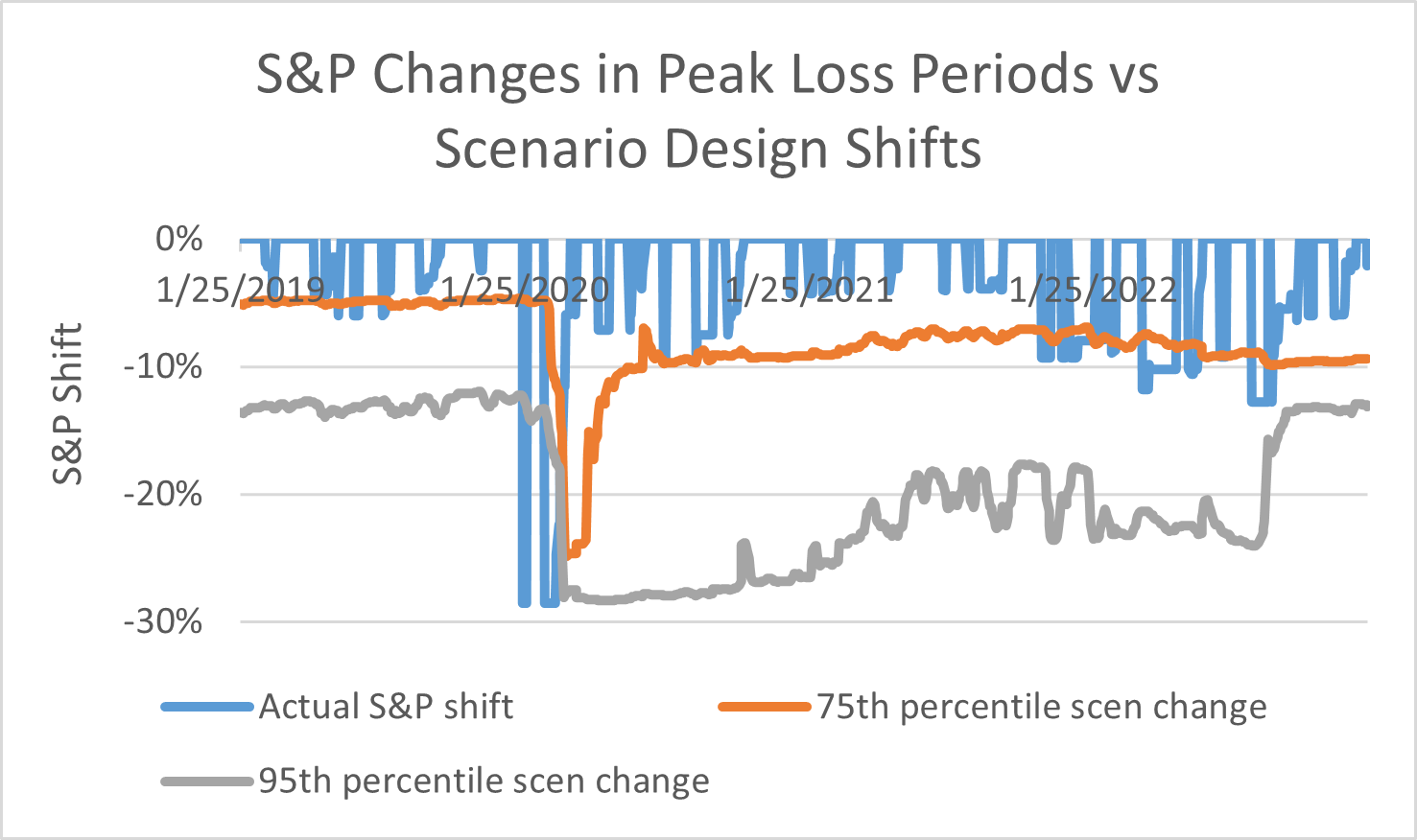}
        \caption{Comparison of S\&P changes in historical peak loss periods vs scenario shocks for the two designed stress scenarios}
  \label{fig:spxshifts}
 \end{figure}

 Figure \ref{fig:spxshifts} shows the S\&P shifts in the designed scenarios vs the S\&P shifts in peak loss periods. Historical S\&P shifts shown in the plot are S\&P shift between start date and end dates of $\text{Loss}^t_{15,45}$ and are zero if a date is not in any $\text{Loss}^t_{15,45}$ time period. One observes that designed S\&P scenario shocks adapt quickly to changing market conditions - increase rapidly in early $2020$ and subside as markets stabilize. 
 
\begin{figure}
    \begin{subfigure}{0.48\textwidth}
        \includegraphics[width=\textwidth]{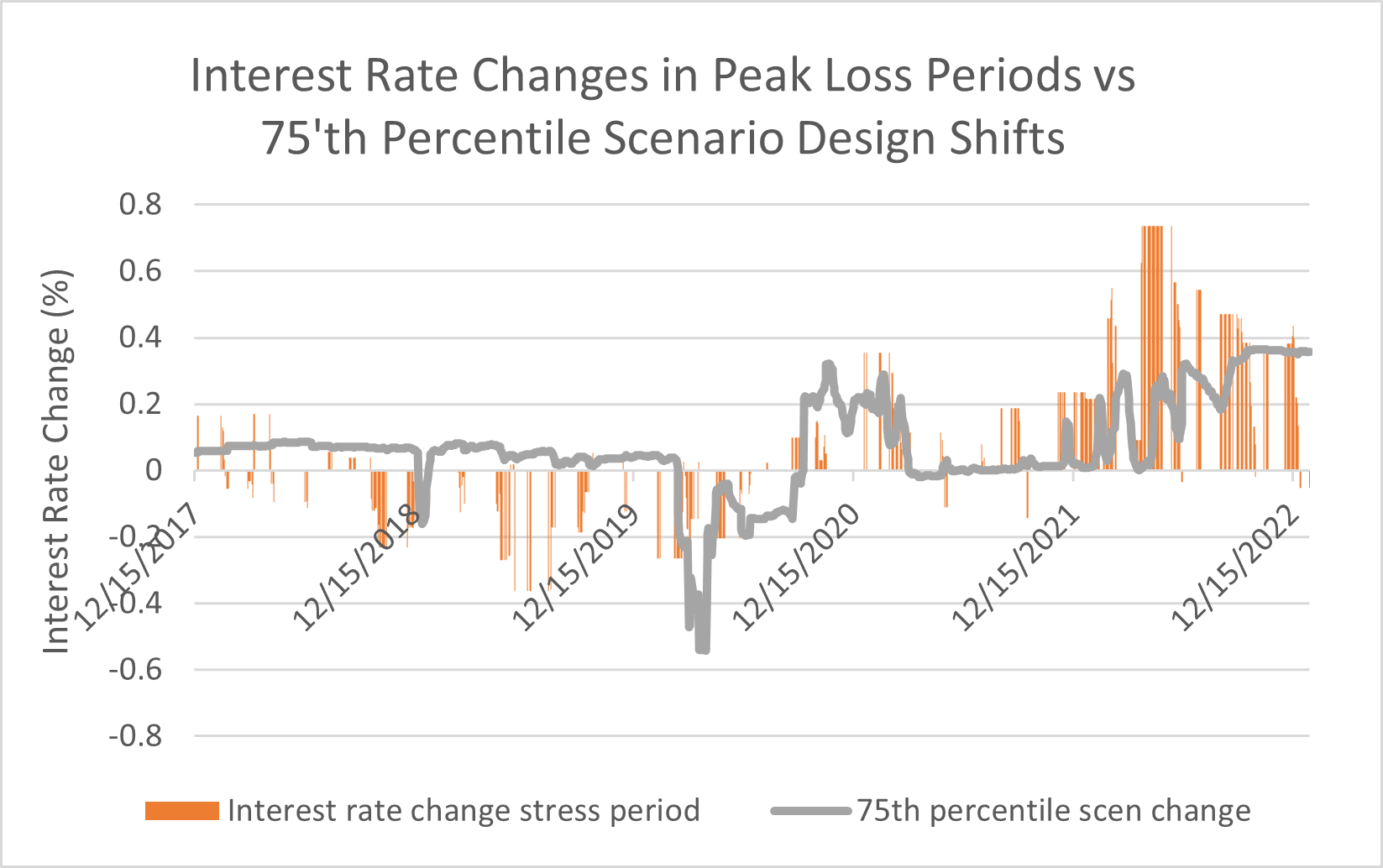}
        \caption{Comparison of ten year Treasury yield changes in historical peak loss periods vs scenario shocks in the stress  scenario with $75\%$ confidence level. Notice that the direction of interest rate changes aligns well with interest rate changes during the stress periods for the portfolio.}
    \end{subfigure} \hfill
    \begin{subfigure}{0.48\textwidth}
        \includegraphics[width=\textwidth]{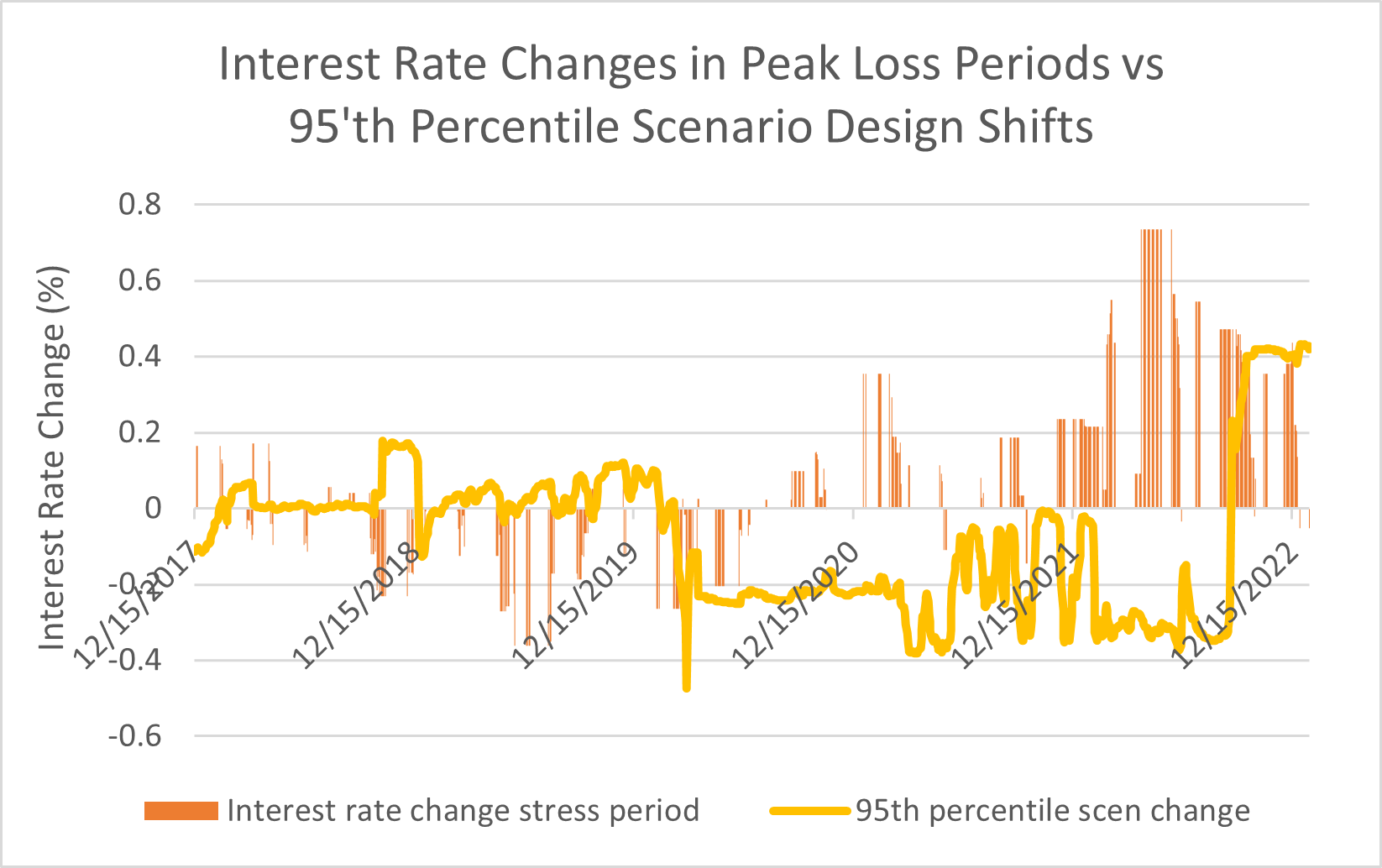}
        \caption{Comparison of ten year Treasury yield changes in historical peak loss periods vs scenario shocks in the stress  scenario with $95\%$ confidence level. For this more stressful scenario, interest rate shocks are usually negative and turn positive only in the year $2022$ after sustained period of Fed tightening.}
    \end{subfigure}
    \caption{Comparison of interest rate changes during stress periods and the stress scenarios designed shifts for those dates. For the less severe scenario (Figure a), interest rate changes from the scenario design are aligned with the changes during stress periods. Initially the shifts are mostly small or negative but after the beginning of Fed tightening, the interest rate shifts in stress scenarios turn positive. For the more severe stress scenario (Figure b), interest rate changes are usually marginally negative. This is because interest rates are unlikely to increase in very stressful market conditions as illustrated in Figure \ref{fig:rateloss}.}
    \label{fig:ratesshifts}
 \end{figure}

\begin{figure}[hbt!]
        \includegraphics[width=\textwidth]{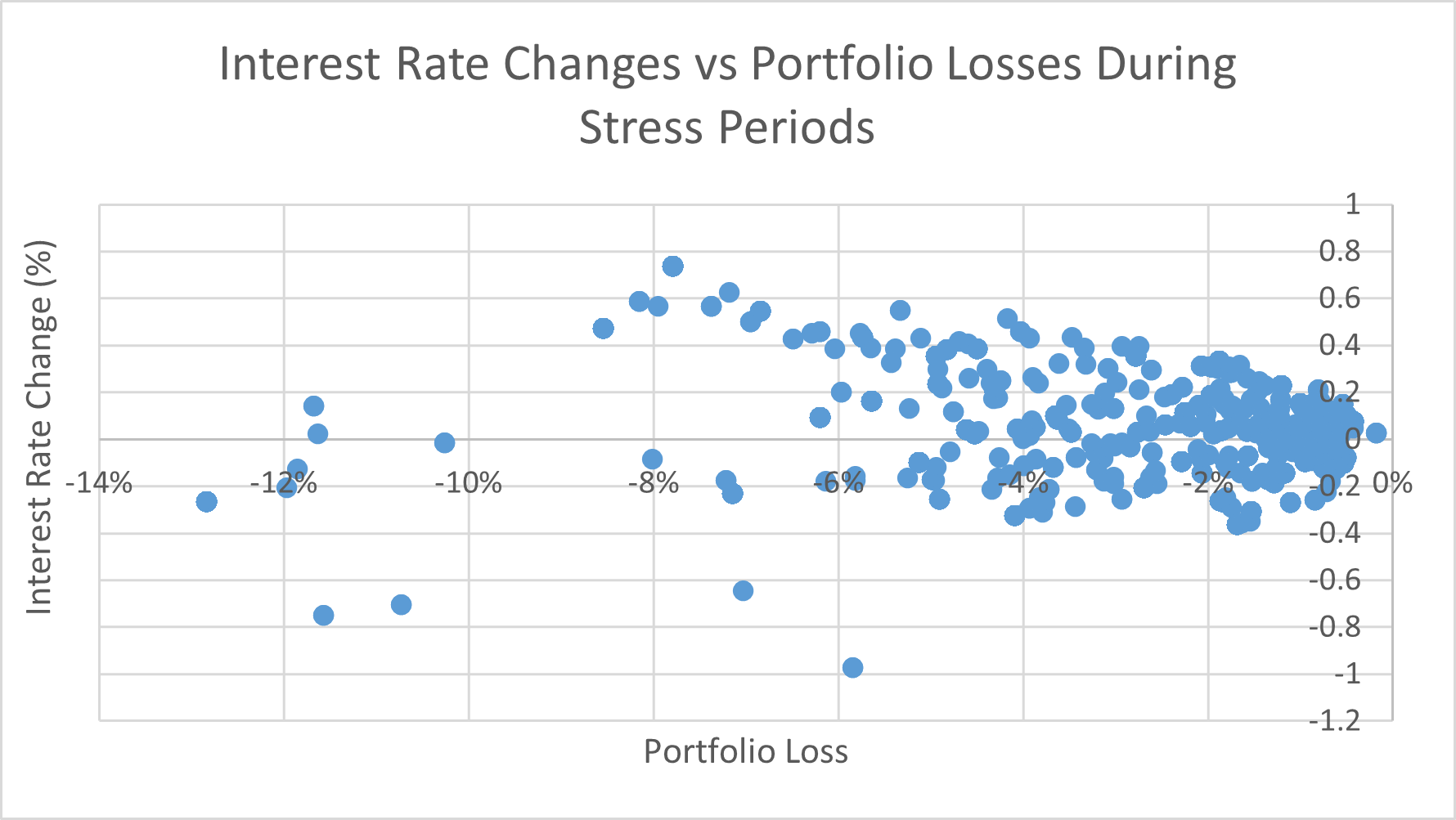}
        \caption{Historical data of interest rate shifts vs. portfolio losses from $2015$ to $2022$. For portfolio losses more severe than $8.5\%$, interest rate changes are rarely positive.}
  \label{fig:rateloss}
 \end{figure}

 Figure \ref{fig:ratesshifts} shows a similar plot of historical interest rate shifts during stress periods vs scenario designed shifts. The period covered in these plots includes both periods of FRB easing monetary conditions during Covid crisis and the subsequent tightening. Interest rate shifts in the less severe stress scenario (Figure \ref{fig:ratesshifts} a) are largely consistent with interest rate changes during the stress periods for the portfolio. However, for the more stressful scenario, interest rate shifts are more likely to be negative (Figure \ref{fig:ratesshifts} b). Thus the scenario shifts can be quite dependent on the severity of the scenario being designed. The fact that the rate shifts are mostly negative in the more stressful scenario design can be understood from the historical data. Figure \ref{fig:rateloss} shows the scatter plot of historical portfolio losses for this portfolio and the corresponding stress periods from the year $2015$ to $2022$. One observes in this plot that for most historical periods when portfolio losses were more than $8.5\%$, interest rate changes were small or negative. Thus historical data suggests that for this portfolio, interest rates are unlikely to increase when the losses are severe (portfolio losses more than $8.5\%$) which is consistent with shifts produced by the severe stress scenario in Figure \ref{fig:ratesshifts} b.

\clearpage
 
\section{Summary}
In this paper, an approach based on variational inference is proposed for designing stress scenarios for portfolios and estimating VaR. The primary objective here is to incorporate prevailing market conditions so that one obtains more realistic and plausible measures of these downside risk measures. In the proposed approach, clusters of market variables (or market regimes) are identified for which the future changes in portfolio values are similar. It is shown that the proposed approach adapts quickly to changes in market conditions. Since one obtains the distribution of portfolio returns or magnitude of losses, one can obtain risk measures for any desired level of confidence. Due to reliance on historical data, the designed stress scenarios are coherent and plausible. 

\section*{Appendix}

We now derive the parameters for the variational distribution based on (\ref{meanfield}). First let us consider the variational density of the cluster mean $\mu_k$. From (\ref{meanfield})
\[
 q_\mu^*(\mu_k; \hat{\mu}_{k}, \hat{R}_{k}) \; \propto \text{exp} \Big( \mathbb{E}_{-\mu_k} \left[ \text{log} \ p(\boldsymbol\mu ,\boldsymbol\theta, \text{\bf{c}}, \text{\bf{d}}, \text{\bf{x}}) \right] \Big)
\]
where $\mathbb{E}_{-\mu_k}$ is expectation with respect to $\{\beta_1, \cdots, \beta_K\}$, $\{c_i \ , \ 1 \leq i \leq T \}$ and $\{\mu_i \ , \ i \neq k.  \}$. From (\ref{muprior}), (\ref{xprior}), (\ref{xcluster}),  and variational parameters for $q_c$ in (\ref{qc1}) one notes that
\begin{align}
    \label{qmu}
    q_\mu^*(\mu_k; \hat{\mu}_{k}, \hat{R}_{k}) & \propto \text{exp} \Big( \mathbb{E}_{-\mu_k} \left[ \text{log} \ p(\boldsymbol\mu ,\boldsymbol\theta, \text{\bf{c}}, \text{\bf{d}}, \text{\bf{x}}) \right] \Big)  \nonumber \\
    & \propto \text{exp} \Big(  \text{log} \ p(\mu_k ; \mu_{k0}, R_{k0} ) +  \sum_{t=1}^T  \mathbb{E}_{c_t} \left[ \text{log} \ p(x_t | c_{t(k)}, \mu_k ) \right] \Big) + \text{const}  \nonumber \\
    & \propto \text{exp} \Big(  -\frac{1}{2}(\mu_k-\mu_{k0})'R_{k0}^{-1}(\mu_k-\mu_{k0}) 
    + \sum_{t=1}^T \mathbb{E}_{c_t} \left[ c_{t(k)}=1 \right] \ \text{log} \ p(x_t | \mu_k ) \Big) + \text{const} \Big) \nonumber \\
    & \propto \text{exp} \Big( \mu_k'R_{k0}^{-1}\mu_{k0} - \frac{1}{2} \mu_k'R_{k0}^{-1}\mu_k
    + \sum_{t=1}^T \phi_{tk} \left[ -\frac{1}{2}(x_t-\mu_{k})'M^{-1}(x_t-\mu_{k}) \right]+ \text{const} \Big)
    \nonumber \\
    & \propto \text{exp} \Big( - \frac{1}{2} (\mu_k -\hat{\mu}_{k})' \hat{R}_{k}^{-1}(\mu_k -\hat{\mu}_{k}) + 
     \text{const} \Big)
\end{align}
where "const" includes constant terms that do not depend on $\mu_k$, $\mathbb{E}_{c_t}$ is expectation with respect to $q_c$ and the variational parameters $\hat{\mu}_{k}$ and $\hat{R}_{k}$ are defined as follows:
\begin{align}
    \label{hatmuR}
    \hat{R}_{k} & := \left[R_{k0}^{-1}  + M^{-1} \sum_{t=1}^T \phi_{tk} \right]^{-1} \nonumber \\
    \hat{\mu}_{k} & := \hat{R}_{k} \left[ R_{k0}^{-1} \mu_{k0} + M^{-1} \sum_{t=1}^T \phi_{tk} x_t       \right]
\end{align}
From the exponential distribution in (\ref{qmu}) one notes that the variational distribution of $\mu_k$, described by $q_\mu^*(\mu_k)$, is Gaussian with mean $\hat{\mu}_{k}$ and variance $\hat{R}_{k}$. Note that in obtaining the above, we did not make any assumptions about the distribution family of $q_\mu^*(\mu_k)$.

We next consider the variational density of $\theta_k$. Recall $d_{t(j)}=1$ is an indicator function which is $1$ if return is in category $j$ at time $t$ and $0$ otherwise. From (\ref{dirich}), (\ref{dprior}), we note the following for any $j \in \{1,\cdots, J\}$:
\begin{align}
    \label{qtheta}
    q_\theta^*(\theta_{k_j}; \hat{\alpha}_{k}) & \propto \text{exp} \Big( \mathbb{E}_{-\theta_k} \left[ \text{log} \ p(\boldsymbol\mu ,\boldsymbol\theta, \text{\bf{c}}, \text{\bf{d}}, \text{\bf{x}}) \right] \Big)  \nonumber \\
    & \propto \text{exp} \Big(  \log \ p(\theta_{k_j} ; \alpha_{k}) +  \sum_{t=1}^T  \mathbb{E}_{\theta_k} \left[ \text{log} \ p(d_{t} | \text{\bf{c}}, \boldsymbol\theta ) + \text{const}  \right] \Big)  \nonumber \\
    & \propto \text{exp} \Big(  (\alpha_{k_j} - 1) \text{log}(\theta_{k_j})
    + \sum_{t=1}^T \mathbb{E}_{c_t} \left[ c_{t(k)}=1 \right] \ \ d_{t(j)} \log (\theta_{k_j}) + \text{const}  \Big)  \nonumber \\
    & \propto \text{exp} \Big(  (\alpha_{k_j} - 1) \text{log}(\theta_{k_j})
    + \sum_{t=1}^T \phi_{tk} \ d_{t(j)}  \text{log} (\theta_{k_j})+ \text{const} \Big)
    \nonumber \\
    & \propto \text{exp} \Big( \{ \alpha_{k_j} - 1+ \sum_{t=1}^T \phi_{tk} \ d_{t(j)} \} \text{log}(\theta_{k_j}) + 
     \text{const} \Big) \nonumber \\
\end{align}
Since for any $k$, $\sum_j \theta_{k_j}=1$, the above implies that $q_\theta^*(\theta_{k})$ is Dirichet distribution of form (\ref{qtheta1}) with the parameters $\hat{\alpha}_{k_j}$ as described in (\ref{alphaest}).

We finally consider the probability for cluster assignment $k$ at time $t$. From (\ref{meanfield}) and (\ref{logP}) one notes that
\begin{align}
    \label{qcluster}
    q_c^*(c_{t(k)}=1; \phi_t) & \propto \text{exp} \Big( \mathbb{E}_{-c_{t}} \left[ \text{log} \ p(\boldsymbol\mu ,\boldsymbol\theta, \text{\bf{c}}, \text{\bf{d}}, \text{\bf{x}}) \right] \Big) \ \ \ ( \text{expectation with respect to all random variables other than } c_t ) \nonumber \\
    & \propto \text{exp} \Big( \mathbb{E}_{-c_{t}} \left[ \text{log} \ p(c_{t(k)}=1,x_t,d_t | \pi, \boldsymbol\mu_, \boldsymbol\theta ) \right]  + \text{const} \Big) \nonumber \\
    & \propto \text{exp} \Big( \mathbb{E} \left[ \text{log} \ p(c_{t(k)}=1 | \pi ) \right] + \mathbb{E}_\mu \left[ \text{log} \ p(x_t | c_{t(k)}=1, \mu_k ) \right] + \mathbb{E}_\theta \left[ \text{log} \ p(d_t | c_{t(k)}=1, \theta_k )
    \right] + \text{const} \Big) \nonumber \\
    & \propto \text{exp} \Big( \mathbb{E} \left[ \text{log} \ p(c_{t(k)}=1 | \pi ) \right] + \mathbb{E}_\mu \left[ \text{log} \ p(x_t | c_{t(k)}=1, \mu_k ) \right] + \mathbb{E}_\theta \left[ \text{log} \ (\sum_{j=1}^J d_{t(j)} \theta_{k_j}) \right]  + \text{const} \Big)
\end{align}
where $\mathbb{E}_{-c_{t}}$ is expectation with respect to $\{\theta_1, \cdots, \theta_K\}$, $\{\mu_1, \cdots, \mu_K \}$ and $\{c_i ; i \neq t \}$ and the expectation of all the terms that do not depend on $c_t(k)$ are lumped into the constant term. $\mathbb{E}_\mu$ and $\mathbb{E}_\theta$ denote expectations with respect to variational distributions $q_\mu(\mu)$ and $q_\theta(\theta)$ respectively. For the last term above, one notes that log$(\sum_{j=1}^J d_{t(j)} \theta_{k_j})=$ log$(\theta_{k_{j^*}})$ if the return at time $t$ is in the $j^*$'th category. The first term is the log prior of $c_{t(k)}$:  
\begin{equation}
    \label{ElogC}
    \mathbb{E} \left[ \text{log} \ p(c_{t(k)}=1 | \pi ) \right] = \text{log} (\pi_k) \ \; \; \ \text{(from (\ref{cluster}))}
\end{equation}
From (\ref{xprior}) and (\ref{xcluster}) the expected log probability of data $x_t$ is:
\begin{align}
\label{Elogx}
\mathbb{E}_{\mu} \left[ \text{log} \ p(x_t | c_{t(k)}=1 ) \right] &  
  =  \mathbb{E}_\mu  \text{log} \ p(x_t | \mu_k ) \nonumber \\
& =  -\frac{n}{2} \text{log} 2 \pi- \frac{1}{2} \text{log} |M| - \frac{1}{2} x_t'M^{-1} x_t + x_t'M^{-1} \ \mathbb{E}_\mu( \mu_k) - \frac{1}{2} \text{trace}(M^{-1} \ \mathbb{E}_\mu (\mu_k \mu_k'))  \nonumber \\
& =   x_t'M^{-1} \ \mathbb{E}_\mu ( \mu_k) - \frac{1}{2} \text{trace}(M^{-1} \ \mathbb{E}_\mu (\mu_k \mu_k'))  + \text{const} 
\nonumber \\
& =  x_t'M^{-1} \hat{\mu}_k - \frac{1}{2} \text{trace}(M^{-1} (\hat{\mu}_k \hat{\mu}_k'+\hat{R}_k))  + \text{const}
\end{align}
where "const" denotes a constant term that does not depend on $c_{t(k)}$. In obtaining the last step above we have used the fact under the variational distribution $q^*_\mu$ described in (\ref{qmu}), $\mu_k$ is normally distributed with mean $\hat{\mu}_k$ and variance $\hat{R}_k$.

From (\ref{qtheta1}), the expected value of log probability of assignment is:
\begin{align}
\label{Elogtheta}
\mathbb{E}_\theta \left[ \text{log} \ (\sum_{j=1}^J d_{t(j)} \theta_{k_j})  \right] & = 
  \Psi (\sum_{j=1}^J d_{t(j)} \hat{\alpha}_{k_j}) - \Psi(\sum_{j=1}^J\hat{\alpha}_{k_j}) 
\end{align}
where $\Psi()$ is the digamma function.  
Combining (\ref{qcluster}), (\ref{ElogC}), (\ref{Elogx}), and (\ref{Elogtheta}), the variational update for the $k'th$ cluster assignment at time $t$ is
\begin{equation}
    q_c^*(c_{t(k)}; \phi_t) \propto  r_{t(k)} \nonumber
\end{equation}
where 
\begin{equation}
    r_{t(k)}:=\text{exp} \left[ \text{log} (\pi_k) + x_t'M^{-1} \hat{\mu}_k - \frac{1}{2} \text{trace}(M^{-1} (\hat{\mu}_k \hat{\mu}_k'+\hat{R}_k))
    + \Psi (\sum_{j=1}^J d_{t(j)} \hat{\alpha}_{k_j}) - \Psi(\sum_{j=1}^J\hat{\alpha}_{k_j})  \right] \nonumber
\end{equation}
Since the sum of probabilities of cluster assignments is one, from the above one observes that
\begin{equation}
    \label{clusterprob}
    q_c^*(c_{t(k)}=1)=\phi_{tk} = \frac{r_{t(k)}}{\sum_{k=1}^K r_{t(k)}}
\end{equation}

\bibliographystyle{unsrt}  

\end{document}